\documentclass[onecolumn]{IEEEtran}
\IEEEoverridecommandlockouts
\usepackage{cite}
\usepackage{amsmath,amssymb,amsfonts}
\usepackage{algorithmic}
\usepackage{graphicx}
\usepackage{textcomp}
\usepackage{xcolor}
\usepackage{url}
\usepackage{epsfig}
\usepackage{booktabs} 
\usepackage{multirow}
\usepackage{epstopdf}
\usepackage{subfig}
\usepackage{graphicx}
\usepackage{xspace}
\usepackage{cite}
\usepackage[T1]{fontenc}
\usepackage{balance}
\usepackage{amsmath} 
\usepackage{amssymb} 
\usepackage[linesnumbered,ruled,vlined]{algorithm2e}

\usepackage{wasysym} 

\usepackage{ragged2e} 
\usepackage{caption}
\renewcommand{\raggedright}{\leftskip=0pt \rightskip=0pt plus 0cm}

\newtheorem{definition}{Definition}

\newtheorem{theorem}{Theorem}
\newtheorem{claim}{Claim}[section]

\newcommand{\ie}{\emph{i.e.,}\xspace}
\newcommand{\eat}[1]{}

\newcommand{\eg}{\emph{e.g.,}\xspace}
\newcommand{\etal}{\emph{et al.}\xspace}

\def\BibTeX{{\rm B\kern-.05em{\sc i\kern-.025em b}\kern-.08em
    T\kern-.1667em\lower.7ex\hbox{E}\kern-.125emX}}
\begin{document}

\title{DISCO: Influence Maximization Meets Network Embedding and Deep Learning}

\author{%
	{Hui Li{\small $^{\#}$}, Mengting Xu{\small $^{\#}$}, Sourav S Bhowmick{\small $^{*}$}, Changsheng Sun{\small $^{\dag}$},  Zhongyuan Jiang{\small $^{\ddag}$}, Jiangtao Cui{\small $^{\dag}$} }%
	\vspace{1.6mm}\\
	\fontsize{10}{10}\selectfont\itshape
	$^{\#}$\, State Key Laboratory of Integrated Services Networks, Xidian University, China\\
	\fontsize{9}{9}\selectfont\ttfamily\upshape
	%
	Emails: hli@xidian.edu.cn, mtingxu@stu.xidian.edu.cn
	\vspace{1.2mm}\\
	\fontsize{10}{10}\selectfont\rmfamily\itshape
	$^{*}$\,School of Computer Science and Engineering, Nanyang Technological University, Singapore\\
	\fontsize{9}{9}\selectfont\ttfamily\upshape
	Email: assourav@ntu.edu.sg
		\vspace{1.2mm}\\
	\fontsize{10}{10}\selectfont\rmfamily\itshape
	$^{\dag}$\,School of Computer Science and Technology, Xidian University, China\\
	\fontsize{9}{9}\selectfont\ttfamily\upshape
	Email: me@casun.xyz, cuijt@xidian.edu.cn
	\vspace{1.2mm}\\
	\fontsize{10}{10}\selectfont\rmfamily\itshape
	$^{\ddag}$\,School of Cyber Engineering, Xidian University, China\\
	\fontsize{9}{9}\selectfont\ttfamily\upshape
	Email: zyjiang@xidian.edu.cn
}


\maketitle

\begin{abstract}
Since its introduction in 2003, the influence maximization (\textsc{im}) problem has drawn significant research attention in the literature. The aim of \textsc{im} is to select a set of $k$ users who can influence the most individuals in the social network. The problem is proven to be NP-hard. A large number of approximate algorithms have been proposed to address this problem. The state-of-the-art algorithms estimate the expected influence of nodes based on sampled diffusion paths. As the number of required samples have been recently proven to be lower bounded by a particular threshold that presets tradeoff between the accuracy and efficiency, the result quality of these traditional solutions is hard to be further improved without sacrificing efficiency. In this paper, we present an orthogonal and novel paradigm to address the \textsc{im} problem by leveraging deep learning models to estimate the expected influence. Specifically, we present a novel framework called \textsc{disco} that incorporates \textit{network embedding} and \textit{deep reinforcement learning} techniques to address this problem. Experimental study on real-world networks demonstrates that \textsc{disco} achieves the best performance w.r.t efficiency and influence spread quality compared to state-of-the-art classical solutions. Besides, we also show that the learning model exhibits good generality.
\end{abstract}


\section{Introduction}\label{sec:intro}
Online social networks have become an important platform for people to communicate, share knowledge and disseminate information. Given the widespread usage of social media, individuals' ideas, preferences and behavior are often influenced by their peers or friends in the social networks that they participate in. Since the last decade, \textit{influence maximization} (\textsc{im}) problem has been extensively adopted to model the diffusion of innovations and ideas. The purpose of \textsc{im} is to select a set of $k$ seed nodes who can influence the most individuals in the network~\cite{Kempe:2003:MSI:956750.956769}. For instance, an advertiser may wish to send promotional material about a product to $k$ users of a social network that are likely to sway the largest number of users to buy the product.

A large number of greedy and heuristic-based \textsc{im} solutions have been proposed in the literature to improve efficiency, scalability, or influence quality. State-of-the-art \textsc{im} techniques attempt to generate $(1-1/e-\epsilon)$-approximate solutions with a smaller number of \textsc{ris} (\textit{Random Interleaved Sampling}) samples, which are mainly used to estimate the expected maximum influence (denoted as $\sigma(v,S)$) for an arbitrary node $v$ given the current selected seeds $S$. They use sophisticated estimation methods to reduce the number of \textsc{ris} samples closer to a theoretical threshold $\theta$ \cite{Tang2015}. As $\theta$ provides a lower bound for the number of required \textsc{ris} samples, these methods have to undertake a \emph{diffusion sampling phase} and generate enough propagation samples in order to estimate the expected influence before selecting a seed. Despite the improvements of the sampling model and reduction in the number of required samples brought by recent studies, the cost of generating these samples is large especially in huge networks. Consequently, \textit{is it possible to avoid the diffusion sampling phase in \textsc{im} solutions?} In this paper, we answer to this question affirmatively by proposing a novel framework that for the first time utilizes network embedding and deep learning to tackle the \textsc{im} problem.

The core challenge in \textsc{im} lies in the process of estimating $\sigma(v,S)$ given $v$ and the partial solution $S$, which is known to be \#P-hard~\cite{Kempe:2003:MSI:956750.956769}. Traditional \textsc{im} solutions address it by sampling the diffusion paths to generate an unbiased estimate for $\sigma(v,S)$. In essence, $\sigma(v,S)$ can be viewed as a mapping $\sigma:V\times G\times \Psi\rightarrow \mathbb{R}$, where $G(V,E)$ denotes the network and $\Psi$ refers to the set of diffusion models, which defines the criteria to determine whether a node is influenced or not. That is, given a network $G$, an arbitrary node $v\in V$, and a particular diffusion model (\eg Independent Cascade, Linear Threshold model~\cite{Kempe:2003:MSI:956750.956769}), $\sigma$ outputs the expected maximum number of influenced nodes by $v$. In this paper, we advocate that it is possible to \textit{predict} the expected influence of $v$ if we can approximate the mapping $\sigma$ as $y=\tilde{\sigma}(v,S;\Theta)$ and \emph{learn} the values of the parameters $\Theta$ by utilizing some machine learning methods. Fortunately, a series of complex mappings have been recently approximated using deep learning techniques~\cite{allen2011acquiring,kim2017co,DBLP:journals/corr/DaiKZDS17}. We leverage on these recent results and present a learning-based framework for the \textsc{im} problem called \textsc{disco} (\textbf{D}eep learn\textbf{I}ng-ba\textbf{S}ed influen\textbf{C}e maximizati\textbf{O}n).

Learning the mapping $\sigma$ is a non-trivial and challenging problem. First, we need to transform the topology information of the target network into features. Second, there is no labeled expected maximum influence to train $\sigma$. Hence, supervised learning approaches cannot be adopted in this scenario. To address these challenges, we \textit{integrate} deep reinforcement learning~\cite{mnih2015human,mnih2013playing} and network embedding techniques~\cite{DBLP:journals/corr/DaiDS16} in \textsc{disco}. To learn the mapping, we adopt a network embedding method to represent the network topology as vector-based features, which further acts as input to deep reinforcement learning. For the learning phase, inspired by recent progress in combinatorial optimization problems, we propose a model to approximate $\sigma(v,S)$ as $\tilde{\sigma}(v,S;\Theta)$ using a deep reinforcement learning technique. For each potential seed, instead of estimating its maximum influence by sampling the diffusion paths, we directly predict its influence via learned mapping function. Moreover, we show that under our model the difference in the predicted influences between any pair of nodes will hardly change whenever a new seed is selected. Therefore, instead of iteratively select $k$ seeds, in our model we are able to select all seeds at the same time. Notably, once a mapping function is learned, it can be applied to multiple homogeneous networks. The main contributions of the paper can be summarized as follows.

\begin{itemize} \itemsep= 0ex
\item We present a novel framework that exploits learning methods to solve the classical \textsc{im} problem. \textit{To the best of our knowledge, this is the first deep learning-based solution to the \textsc{im} problem.}
\item We present a novel learning method to approximate $\sigma(v,S)$ as $\tilde{\sigma}(v,S;\Theta)$, by exploiting deep reinforcement learning and network embedding.

\item Our proposed framework generates seed sets with superior running time to state-of-the-art machine learning-oblivious \textsc{im} techniques without compromising on result quality. Specifically, it is up to 36 times faster than \textsc{ssa}~\cite{DBLP:journals/corr/NguyenTD16}. Furthermore, the influence quality of our method is slightly better than the traditional methods.
\item We show how \textsc{disco} can be utilized to address the \textsc{im} problem in large-scale networks or in the presence of evolutionary networks.
\end{itemize}

The rest of this paper is organized as follows. Section~\ref{sec:relwork} reviews related work. We formally present the learning-based \textsc{im} problem in Section~\ref{sec:problem}. We introduce the training and testing procedures in our framework in Sections~\ref{sec:learning} and~\ref{sec:seedsel}, respectively. Experimental results are reported in Section~\ref{sec:exp}. Finally, we conclude this work in Section~\ref{sec:concl}.

\vspace{-0ex}\section{Related Work}\label{sec:relwork}
In this section, we review research on influence maximization and network embedding.
\vspace{-0ex} \subsection{Influence Maximization}
Since the elegant work in \cite{Kempe:2003:MSI:956750.956769}, the \textsc{im} problem has been studied extensively. Kempe \etal~\cite{Kempe:2003:MSI:956750.956769} proved that the problem of \textsc{im} is NP-hard and provided a greedy algorithm that produces a $(1-1/e-\epsilon)$-approximate solution. Later, \textsc{celf} algorithm~\cite{Leskovec:2007:COD:1281192.1281239} is proposed, which reduces running time through a lazy update strategy. It is nearly 700 times faster than the hill-climb algorithm proposed by~\cite{Kempe:2003:MSI:956750.956769}. \textsc{celf}++ further optimizes \textsc{celf} by exploiting the submodularity property \cite{Goyal:2011:COG:1963192.1963217}. Although \textsc{celf}++ maintains more information than \textsc{celf}, it has nearly 7\% efficiency improvement over \textsc{celf}. However, it is still time consuming for these approximate algorithms.

Besides these approximate algorithms, many excellent heuristic algorithms~\cite{Chen:2009:EIM:1557019.1557047,Jiang:2011:SAB:2900423.2900443,Goyal:2011:SEA:2117684.2118230}, which do not provide an approximation ratio, have been proposed to reduce running time further. Compared to \textsc{celf}++/\textsc{celf}, these heuristic algorithms have improved the efficiency by at least two orders of magnitude. Although these heuristic algorithms greatly reduce the execution time, they sacrifice accuracy of the results significantly~\cite{Chen:2009:EIM:1557019.1557047,Jiang:2011:SAB:2900423.2900443,Goyal:2011:SEA:2117684.2118230}.

Recently, Borgs et al.~\cite{Borgs:2014:MSI:2634074.2634144} proposed a new \textsc{im} sampling method called \textit{reverse influence sampling} (\textsc{ris}). It sets the threshold to determine how many \textit{reverse reachable sets} are generated from the network and then selects the node that covers the most number of these sets. Based on this idea, a series of advanced approximate algorithms have been proposed. These algorithms can not only provide an approximation ratio, but also exhibit competitive running time compared to heuristic algorithms. \textsc{tim}/\textsc{tim}+~\cite{DBLP:journals/corr/TangXS14} algorithm significantly improved the efficiency of~\cite{Borgs:2014:MSI:2634074.2634144} and is the first \textsc{ris}-based approximate algorithm to achieve competitive efficiency with heuristic algorithms.
Subsequently, \textsc{imm}~\cite{Tang2015} utilized the notion of martingales to significantly reduce computation time in practice while retaining \textsc{tim}'s approximation guarantee. Specifically, it can return a $(1-1/e-\epsilon)$-approximate solution with at least $1-1/n^\ell$ probability in $O((k+\ell)(n+m)log\ n/\epsilon^2$ expected time.

More recently, \textsc{ssa} and \textsc{d-ssa}~\cite{DBLP:journals/corr/NguyenTD16} claimed to be superior to \textsc{imm} w.r.t. running time while sharing the same accuracy. \textsc{ssa} is based on the \textit{Stop-and-Stare} strategy where the number of required \textsc{ris} samples can be closer to the theoretical lower bound. However, Huang et al.~\cite{Huang:2017:RSA:3099622.3099623} performed a rigorous theoretical and experimental analysis of these algorithms and found that the results reported in~\cite{DBLP:journals/corr/NguyenTD16} are somewhat unfair.


In recent years, there are increasing efforts to address \textsc{im} problems with the help of learning methods. \cite{DBLP:conf/webi/AliWC18,DBLP:conf/kdd/LinLC15} have both employed reinforcement learning model to find best strategy in competitive \textsc{im} problem, which was thoroughly studied in~\cite{carnes2007maximizing,li2015getreal}. Different from \cite{DBLP:conf/webi/AliWC18}, the authors in \cite{DBLP:conf/kdd/LinLC15} addressed the competitive \textsc{im} in a multi-round and adaptive setting~\cite{DBLP:conf/kdd/SunHYC18,DBLP:journals/pvldb/HanHXTST18}. Both of these works treat the competition between multiple parties as a game; and employ reinforcement learning to find the best policy (\ie strategy) that can maximize the profit given the opponents' choices. They do not natively model the influence estimation as a reinforcement learning problem or node influence as a learning task. Hence, these works are only applicable to competitive setting and are orthogonal to our problem. Besides, \cite{DBLP:conf/icml/VaswaniKWGLS17,DBLP:conf/pkdd/CaoWHW11} adopt learning methods to study the diffusion model and optimize the linear threshold model parameters, respectively. They do not consider \textsc{im} as a machine learning problem and are also orthogonal to our problem.

In summary, \textsc{imm} and \textsc{ssa} are currently recognized as the state-of-the-art methods to solve the \textsc{im} problem according to a benchmarking study~\cite{DBLP:conf/sigmod/AroraGR17}. Notably, none of these efforts integrate machine learning with the \textsc{im} problem to further enhance the performance.

\vspace{-0ex} \subsection{Network Embedding}
Network embedding is to find a mapping function that converts each node in a network into a vector-based representation. The learned vector-based representation can be used as a feature of various tasks based on graphs, such as classification, clustering, link prediction and visualization. One of the major benefit of network embedding is that the resulting vector representation can be directly feed into most machine learning models to solve specific problems.

Early methods for learning node representations focused primarily on matrix decomposition, which was directly inspired by classical techniques for dimensionality reduction \cite{belkin2002laplacian}. However, these methods introduce a lot of computational cost. Recent approaches aim to learn the embedding of nodes based on random walks. DeepWalk \cite{perozzi2014deepwalk} was proposed as the first network embedding method using deep learning  technology, which compensates for the gap between language modeling and network modeling by treating nodes as words and generating short random walks. LINE \cite{DBLP:journals/corr/abs-1710-09599} uses the Breadth First Search strategy to generate context nodes, in which only nodes that are up to two hops from a given node are considered to be neighbors. In addition, it uses negative sampling to optimize the Skip-gram model compared to the layered softmax used in DeepWalk. node2vec \cite{grover2016node2vec} is a sequence extraction strategy optimized for random walks on DeepWalk framework. It introduces a biased random walk program that combines Breadth First Search and Depth First Search during neighborhood exploration. SDNE \cite{DBLP:conf/kdd/WangC016} captures the nonlinear dependency between nodes by maintaining the proximity between 2-hop neighbors through a deep autoencoder. It designs an objective function that describes both local and global network information, using a semi-supervised approach to fit optimization. It further maintains the proximity between adjacent nodes by minimizing the Euclidean distance between their representations. There is also a kernel-based approach where the feature vectors of the graph come from various graphics kernels\cite{shervashidze2011weisfeiler}. Structure2vec \cite{DBLP:journals/corr/DaiDS16} models each structured data point as a latent variable model, then embeds the graphical model into the feature space, and uses the inner product in the embedded space to define the kernel, which can monitor the learning of the graphical structure. DeepInf~\cite{DBLP:conf/kdd/QiuTMDW018} presents an end-to-end model to learn the probability of a user's action status conditioned on her local neighborhood.

\vspace{-0ex}\section{Problem Statement and Diffusion Model}\label{sec:problem}
In this section, we first formally present the learning-based \textsc{im} problem. Next, we briefly describe the information diffusion models discussed in these definitions.
\vspace{-0ex}\subsection{Problem Statement}
Let $G = (V, E, W)$ be a social network, where $V$ is a set of nodes, $E$ is a set of edges, $|E|=m$, and $|V|=n$. $(u, v) \in E$ represents an edge from node $u$ to node $v$. Let $W$ denote the weight of an edge indicating the strength of the influence. Accordingly, the \textsc{im} problem can be formally defined as follows.

\begin{definition}\label{pro:str}{\em
{\bf (Influence Maximization)} Given a social network $G = (V, E, W)$, an information diffusion model $\psi$, integer $k$, the \textbf{influence maximization problem} aims to select $k$ nodes as the seed set $S$ ($S\subseteq V$), such that, under the diffusion model $\psi$, the expected number of influenced nodes by $S$, namely $\sigma (S)$, is maximized. The problem can be modeled as the following.
\[argmax_{|S|=k,S\subseteq V}{\sigma(S)}\] \/}
\end{definition}

As \textsc{im} is proved to be NP-hard~\cite{Kempe:2003:MSI:956750.956769}, all approximate solutions need to greedily select the next seed with the maximum marginal improvement in expected influence. In particular, let $S_i$ be a partial solution with $i$ seeds (\ie $i\le k$), then in $(i+1)$-th iteration, an approximate algorithm shall choose a node $v$, such that $\sigma(S_{i+1})-\sigma(S_i)$ is maximized, where $S_{i+1}=S_i\bigcup \{v\}$. To facilitate the following discussions, we refer to $\sigma(v,S)$ as the \textit{maximum marginal expected influence} of $v$ given a partial solution $S$. As $\sigma(v,S)$ is \#P-hard to calculate based on $v$ and $S$, traditional efforts in \textsc{im} generate unbiased estimates for $\sigma(v,S)$ using a set of \textsc{ris} samples.
In this paper, we solve the \textsc{im} problem by adopting a completely new strategy \ie a learning method. In our solution, $\sigma(v,S)$ is not estimated using \textsc{ris}-based samplings. Instead, it is approximated using deep learning models. In this regard, we present the definition of \textit{learning-based \textsc{im} problem} as follows.
\begin{definition}\label{pro:str2}{\em
	{\bf (Learning-based Influence Maximization)} [Learning phase.] Given a series of homogeneous networks $\mathcal{G}=\{G_1,\ldots,G_\ell\}$, an information diffusion model $\psi$, train a group of parameters $\Theta$ such that function $y=\tilde{\sigma}(v,S;\Theta)$ can be used to approximate $\sigma(v,S)$ as accurately as possible. [Seeds selection phase.] Given a target network $G$, integer $k$ and a function $y=\tilde{\sigma}(v,S;\Theta)$ that approximately calculates the marginal influence of $v$ with respect to the partial solution $S$, solve the \textsc{im} problem in $G$ with respect to budget $k$ and diffusion model $\psi$.\/}
\end{definition}

\vspace{-0ex}\subsection{Diffusion Models}
Based on the definition of \textsc{im}, one can observe that a diffusion model $\psi$ is vital for the selection of seeds. Currently, there exist two popular diffusion models, namely \textit{Linear Threshold } (\textsc{lt}) and \textit{Independent Cascade} (\textsc{ic}). Throughout a diffusion process, a node has only two states, activated and inactivated. Both models assume that, when a node is activated, its state will not change further.

\textbf{Linear Threshold (\textsc{lt}) model.} The \textsc{lt} model is a special case of the triggering model \cite{kempe2005influential}. To understand the concept, we introduce two related notions. $N(v)$: set of neighbors of node $v$; $N^\eta(v)$: set of activated neighbors of node $v$.

In the \textsc{lt} model, each node $v$ in a graph has a threshold $\theta_v$. For each node $u\in N(v)$, the edge $(v,u)$ has a non-negative weight $w(v,u)\le 1$. Given a graph $G$ and a seed set $S$, and the threshold for each node, this model first activates the nodes in $S$. Then it starts spreading in discrete timestamps according to the following random rule. In each step, an inactivate node $u$ will be activated if $\sum _{u\in N^\eta(v)}w(v,u) \ge \theta _v.$ The newly activated node will attempt to activate its neighbors. The process stops when no more nodes are activated.

\textbf{Independent Cascade (\textsc{ic}) model.} Given a graph $G$ and a seed set $S\subset V$, this model first activates the nodes in $S$, and then starts spreading in discrete timestamps according to the following rule. When a node $u$ is first activated at timestamp $t$, it gets a chance to activate a node in its neighborhood that is not activated. The success probability of activation is $w(v,u)$, and the failure probability is $(1-w(v,u))$. If $v$ is activated successfully, $v$ will become an active node in step $t+1$ and it can no longer activate other nodes in subsequent steps. This process continues until no new nodes can be activated. In other words, whether $u$ can activate $v$ is not affected by previous propagation.

\begin{figure*}[t]
  \centering
  \includegraphics[clip=true,width=0.98\linewidth]{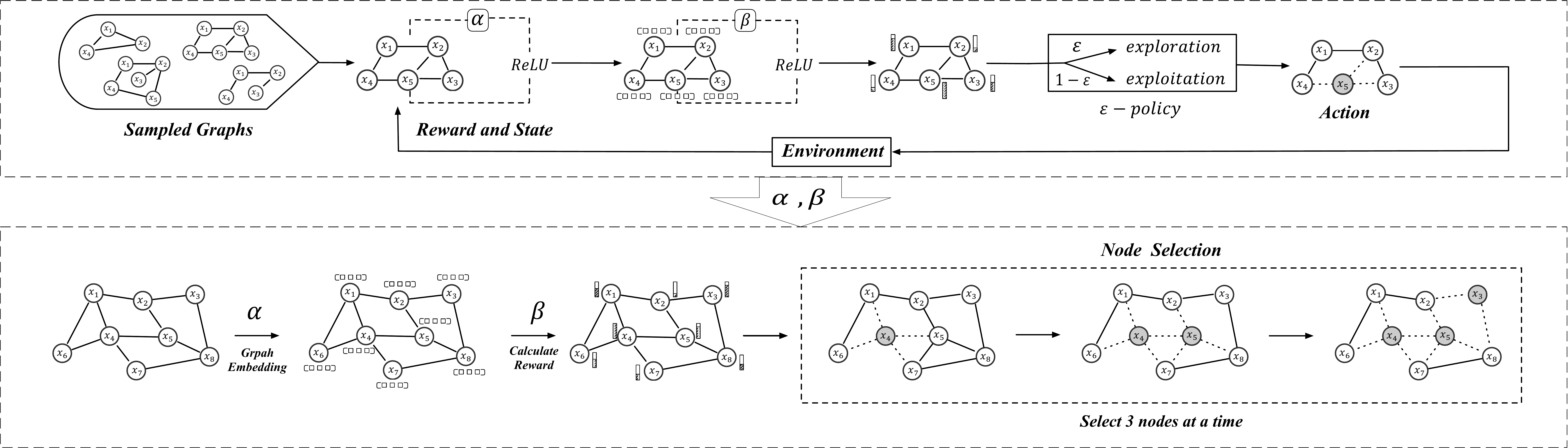}
  \vspace{0ex}\caption{\textsc{disco} framework by incorporating network embedding and deep reinforcement learning. (with $k=3$)}\label{fig:meth}\vspace{-1ex}
\end{figure*}

\vspace{-0ex}\section{Learning the Mapping Function}\label{sec:learning}
Notably, in traditional approximate \textsc{im} solutions, it is inevitable to undertake a diffusion sampling phase to generate a group of \textsc{ris} and the number of required \textsc{ris} is \emph{at least} as large as the threshold $\theta$. In this paper, we turn to deep learning models to avoid the traditional diffusion sampling phase in seeds selection. In this section, we shall present our \textsc{disco} framework that addresses the \textsc{im} problem using deep learning techniques. As remarked earlier, the key challenge in addressing \textsc{im} lies in the estimation of expected maximum influence function $\sigma(v,S)$. A learning-based \textsc{im} solution should inevitably train a model, namely $\tilde{\sigma}(v,S;\Theta)$, that can approximate the mapping $\sigma$ as accurately as possible. Unfortunately, as the influence maximization problem is NP-hard, the ground-truth label for $\sigma$ is hard to acquire. Consequently, we adopt \emph{Deep Q-Network} (\textsc{dqn})~\cite{mnih2015human,mnih2013playing}, a popular deep reinforcement learning model to learn the parameters $\Theta$. In this way, an optimal approximation of $\sigma(v,S)$ can be acquired, \ie $\tilde{\sigma}(v,S;\Theta)$. \eat{Besides, we theoretically show that, under the learned mapping $\tilde{\sigma}$, there is limited difference in the ranking of predicted nodes' influences no matter we update the embedding iteratively or not. Therefore, in the seeds selection phase (\ie test phase in the learning framework), given a trained $\tilde{\sigma}$, we can select the $k$ nodes with the maximum predicted influence without iteratively evaluate the marginal influence. In this way, this selection procedure will construct a seed set in real-time. }

Next, we introduce the learning phase of our framework, which consists of network embedding, training of the parameters $\Theta$, and approximating $\sigma$ with $\tilde{\sigma}$ for use in \textsc{dqn}. The test phase for selecting nodes according to the learned parameters and predicted influences will be detailed in the next section. An overview of the \textsc{disco} framework is depicted in Figure~\ref{fig:meth}, where the top half depicts the learning phase while the bottom half illustrates the test phase (\ie seeds selection).

\vspace{-0ex}\subsection{Embedding the Nodes}
Before \textsc{dqn} model can be applied, we shall first extract features of nodes based on the topological information. To this end, we need the embedding of each node $v\in V$ as a vector $x_v$. Among series of embedding methods, \eg DeepWalk~\cite{perozzi2014deepwalk}, node2vec~\cite{grover2016node2vec}, DeepInf~\cite{DBLP:conf/kdd/QiuTMDW018}, etc., we select Structure2vec~\cite{DBLP:journals/corr/DaiDS16} to accomplish this step due to the following reasons. Firstly, the other alternatives are `transductive' embedding methods, that is, embeddings extracted across graphs are not consistent, since they only care about intra-graph proximity. In our framework, we aim to use a method to complete the cross-graph extraction of embedded results. This means that the parameters trained in the subgraphs can be applied to the target graph. Secondly, since they are unsupervised network embedding methods (\ie DeepWalk and node2vec) or supervised for a particular task (\ie DeepInf), they may not capture the desired information (\ie expected influence spread) within \textsc{im} problem. In our case, the network embedding is trained end-to-end for the optimization target, thus it can be more discriminative.

In order to fully encode the topological context of the nodes in a network and meet the needs of our framework, we present a variation of Structure2vec~\cite{DBLP:journals/corr/DaiDS16} to achieve this task. Structure2vec learns feature space with discriminative information, and embeds latent variable models into feature spaces. It combines the characteristics of nodes, edges and network structure in the current network. These characteristics will recursively aggregate according to the state of the current network. Structure2vec can achieve end-to-end learning through the combination with \textsc{dqn}, and the parameters learned from training graphs are well applied to the test graph. However, the original Structure2vec model cannot be directly applied in \textsc{dqn}. Hence, we present a variant of this as follows.

First, we initialize the vectors of all nodes and set each of them as a $q$-dimensional zero vector\footnote{In line with~\cite{DBLP:journals/corr/DaiDS16}, $q$ is generally set to 64, it can be adjusted according to the size of the network}. After $I$ iterations, ($I$ is usually small, set to 4 or less), each node $v$ reaches to the final state. In the \textsc{im} scenario, we adopt $a_v$ to indicate whether node $v$ is in the partial solution $S$. That is, $a_v=1$ if the node appears in the seed set $S$, else $a_v=0$.  The formula for the update of vectors is as follows:
\begin{equation}\label{eq1}
\small
	x^{(i)} _v:= ReLU(\alpha _1\sum _{u\in N(v)} x^{(i-1)} _u \\
	+\alpha _2\sum _{u\in N(v)} ReLU(\alpha _3 w(v,u))+\alpha _4 a_v).
\end{equation}
In Eq.~\ref{eq1}, $ReLU$ refers to the Rectified Linear Unit of a neural network, $N(v)$ is the neighbor set of node $v$, $x^{(i)}_v$ represents the vector of node $v$ during the $i$-th iteration,  $w(v,u)$ is the weight of edge $(v,u)$, and $\alpha_1,\ldots,\alpha_4$ are the parameters that need to be trained. Although all the neighbors of each node $v$ remain unchanged during the update process, in order to better model the nonlinear mapping of the input space, we add two parameters $\alpha_2$ and $\alpha_3$ to construct two independent layers of Multi-Layer Perceptron in the above formula.

It can be seen that the first two items in the equation aggregate the surrounding information by summing up the neighbors of $v$. Besides, during the iterations, the update formula can also pass information and network characteristics of a node across iterations. When the embeddings of all nodes have been updated, the next iteration begins. After $I$ iterations, the vector of $v$ will contain information about all neighbors within the $I$-hop neighborhood.

After generating the vector of each node in the network, we define the evaluation function $Q$ that will be used in \textsc{dqn}. As \textsc{dqn} employs $Q$ function to evaluate the quality (\ie influence) of each candidate solution (\ie node) at a particular state (\ie partial seeds), it naturally points to $\tilde{\sigma}(v,S;\Theta)$. In this regard, we shall refer to $\tilde{\sigma}(v,S;\Theta)$ as $Q$ function. Traditionally, the $Q$ function is manually set based on experience, which is challenging. Hence, we use neural networks to find the best performing evaluation function. In state $S$, the evaluation function $Q(v,S,\Theta)$ of node $v$ is defined as follows:
\begin{equation}\label{eq2}\small
	Q(v,S,\Theta):= {\beta _1}^TReLU ([\beta _2\sum _{u\in V}x^{I} _u ,\beta _3x^{I} _v]).
	\vspace{-1ex}
\end{equation}
where $x^{I} _v$ is the vector generated after $I$ iterations; $[,]$ is the concatenation operator. Because $ Q(v,S,\Theta)$ is mainly determined by the embedding of the current node $v$ and its surrounding neighbors, the $Q$ function is related to the parameters $\alpha _1 \sim \alpha _4, \beta _1 \sim \beta _3$, all of which need to be learned. We denote all these parameters using $\Theta$ for simplicity. By using matrix operations, we can reduce the input dimension and output a single-valued $Q$ using the idea of value function approximation. After adding a new node $v$ to the seed set, $a_v$ is changed from 0 to 1. In the new state, embedding needs to be performed again, and the $Q$ value of the remaining nodes needs to be recalculated.

When we evaluate the quality of each node, the node with the best marginal expected influence is added to the seed set $S$ by the greedy algorithm until the number of seeds reaches $k$. In general, given a network $G$, a positive integer $k$, a seed set $S$, $\bar{S}= V\backslash S$ as a set of candidate nodes, we shall calculate $Q$ of each node in the candidate set. If $Q(v^\prime,S,\Theta) =argmax_{v\in \bar{S}} Q(v,S,\Theta)$, then we add node $v^\prime$ to $S$.

\vspace{-0ex}\subsection{Model Training via DQN}\label{ssec:modeltrain}
In the \textsc{im} problem, we are unable to acquire sufficient labeled samples for training the parameters $\Theta$ as the exact evaluation of $\sigma(v,S)$ is \#P-hard. Hence, we turn to deep reinforcement learning. Intuitively, deep reinforcement learning enables end-to-end learning from Perception to Action and automatically extracts complex features. The most representative of these approaches is the \textsc{dqn} (Deep Q-Network) algorithm \cite{mnih2015human,mnih2013playing}. \textsc{dqn} exhibits several advantages over existing algorithms. First of all, it uses reward to construct labels through Q-Learning. Second, the correlation between data samples is eliminated by experience replay. Finally, the input to \textsc{dqn} is a graph, and the output is the action and the corresponding $Q$ function. We refer to an evaluation function $Q$ with weights as a \textit{Q-network}. We will now show how to train the Q-network in our scenario.

Reinforcement learning needs to consider the interaction between \textit{Agent} and \textit{Environment}. For our \textsc{im} scenario, we refer to \textit{Agent} as an object with behavior. No matter what kind of scenario it is applied, an \textit{Agent} contains a series of \textit{Action}, \textit{State}, and \textit{Reward}. For the \textsc{im} problem, we define these reinforcement learning components as follows:
\begin{itemize} \itemsep= .5ex
  \item \textit{Action}: $A$ represents actions of the agent. Each action adds a node $v$ ($v \in \bar{S}$) to the current seed set $S$. We use the network embedding for $v$ to denote an action.
  \item \textit{State}: $S$ is the state of the world that an agent can perceive. In the \textsc{im} problem, we use the current seed set to represent the state $S$. The final state $S_k$ is the $k$ nodes that have been selected. The state $S$ is denoted by the embedding of the currently selected node.
  \item \textit{Reward}: Reward $R$ is a real value. Each time the Agent interacts with the environment, the environment returns a reward for evaluating the action, which represent reward or punishment. The state changes from $S$ to $S^\prime$, after adding node $v$ to the current seed set. The increment of the influence range is the reward of $v$ node in state $S$, $R(S,v)=\sigma (S^\prime)-\sigma (S)$. It should be noted that the ultimate goal of learning is to maximize the cumulative reward $\sum ^k _{i=1} R(S_i,v_i)$.
  \item \textit{Transition}: When a node is selected to join the seed set, $a_v$ will change from 0 to 1. We define this process as $P$.
\end{itemize}

Based on these definitions, \textit{Policy} of reinforcement learning can be constructed. \textit{Policy} is the behavior function of the agent, which is a mapping from state to action. It tells the agent how to pick the next action. A policy that picks an optimal action based on the current $Q$ value is referred to as greedy policy. That is, $argmax_{v\in \bar{S}} Q(v,S,\Theta)$. In this paper ,we use $\varepsilon$-policy, which is a strategy including a random policy and a greedy policy. The advantage of $\varepsilon$-policy is as follows. The usage of random policy can expand the search range, and a greedy policy can facilitate refinement of the $Q$ value. Generally, $\varepsilon$ is a small value as the probability of selecting random actions. The $\varepsilon$-policy balances exploration and exploitation by adapting $\varepsilon$.

To train the model, we also need to define a loss function.
In order to provide a labeled sample to the $Q$ network, \textsc{dqn} adds a target $Q$ function: $R+\gamma max_vQ(v^\prime,S_{i+1},\Theta)$ compared to Q-learning, $\gamma$ is decaying factor, which controls the importance of future rewards in learning. If $\gamma=0$, the model will not learn any future reward information, become short-sighted, and only pay attention to the current interests; if $\gamma>=1$, the expected value is continuously accumulated, there is no attenuation, so the expected value may diverge. Therefore, $\gamma$ is generally set to a number slightly smaller than 1.

In the \textsc{im} problem, the reward of an action can only be calculated accurately after a series of actions. Therefore, we use $\delta$-step Q-learning, which can effectively handle delayed rewards. Specifically, we wait $\delta$ steps before updating the parameters to collect rewards for the future more accurately. We set target as $\sum ^{\delta-1} _{i=0}R_i+\gamma max_vQ(v^\prime,S_{i+1},\Theta)$. Hence, the loss function in the neural network is:
\begin{equation}\small
	L(\Theta)=\mathbb{E} [(\sum ^{\delta-1} _{i=0}R_i+\gamma max_vQ(v^\prime,S_{i+1},\Theta)
	-Q(v_{i},S_{i},\Theta))^2]
\end{equation}
\begin{algorithm}[t]
	\caption{Parameter inferencing}\label{alg:train1}
	\KwIn{Batch of training network, a positive number $k$,experience replay memory $\mathcal{J}$ to capacity $N$.}
	\KwOut{parameters $\Theta=\{\alpha_1,\ldots,\alpha_4,\beta_1,\ldots,\beta_3\}$ }
	\For{episode $e = 0; e<E $}{
		Given a networks $G$ and set seed set $S =\phi $\;
		\For{$i=0$ to $k-1$}{
			Compute the embedding for each node $v \in V$ \;
			Calculate the $Q$ value of each node\;
			$v_i =
			\begin{cases}
			\mbox{Random node } v \in \bar{S},   & \mbox{w.p. } \varepsilon  \\
			argmax_{v\in \bar{S}} Q(v, S, \Theta), &  \mbox{w.p. } 1-\varepsilon
			\end{cases}$\;
			add $v_i$ to $S$\;
			update $S,\bar{S}$\;
			\If{$i \ge \delta$}{
				Add tuple $\langle S_i, v_i, \sum ^{i+\delta} _i R_i, S_{i+\delta}\rangle $ to $\mathcal{J}$\;
				Sample random batch from $j \sim \mathcal{J}$\;
				Update $\Theta$ by SGD with $j$\;
			}
		}
	}
\end{algorithm}
The training process is outlined in Algorithm~\ref{alg:train1}. Because of the continuity between the data, the training is easily affected by the sample distribution and it is difficult to converge. In this regard, we use experience replay to update the parameters. In this way, the training samples stored in the dataset can be randomly extracted, so that the training process becomes smooth. We use a batch of homogeneous networks during the training. The episode represents the process of obtaining a complete sequence of a network seed set $S$ (Lines 1-11). For each network that is trained, the seed set $S$ is initialized (Lines 2). Next, we embed each node and calculate its $Q$ value. According to the embedding process discussed earlier in this section, we update the nodes' embeddings and calculate the $Q$ value for each node (Lines 3-4). After getting the influence quality of each node, we apply the aforementioned $\varepsilon$-policy. In particular, we randomly select a node with probability $\varepsilon$; otherwise we select the node with the highest $Q$ value with respect to current partial solution. The selected node is then added to the seed set (Lines 5-7). Finally, in order to accurately collect rewards for the future, we will wait for $\delta$ steps before updating the parameters. After $\delta$ steps, the empirical sample $j$ randomly extracted from replay memory $\mathcal {J}$ is adopted to update the parameters using \textsc{sgd} (Stochastic Gradient Descent) (Lines 9-12).

\eat{Notably, the complete pseudocode and implementation details for this part is outlined in Appendix~\ref{app:code}.}
\vspace{-0ex}\section{Seeds Selection}\label{sec:seedsel}
\begin{algorithm}[t]
	\caption{Seeds selection process}\label{alg:test}
	\KwIn{network $G =(V,E,W)$, a positive number $k$, parameters $\Theta=\{\alpha_1,\ldots,\alpha_4,\beta_1,\ldots,\beta_3\}$}
	\KwOut{optimal solution $S_k$ }
	Initialize a seed set $S =\phi $ and $x^{(0)} _v =0 (v \in V)$.\\
	\For{$j=1$ to $I$}{
		\For{$v \in V$}{
			$x^{(j)} _v:= ReLU(\alpha _1\sum _{u\in N(v)} x^{(j-1)} _u +\alpha _2\sum _{u\in N(v)} ReLU(\alpha _3 w(v,u))+\alpha _4 a_v)$
		}
	}
		\For{$v \in V$}{
			$Q(v,\emptyset,\Theta )={\beta_1}^TReLU ([\beta _2\sum _{u\in V}x^{I} _u,\beta _3x^{I} _v])$
		}
	$S_k\leftarrow$top-$k$ node $v$ with the highest $Q(v,\emptyset,\Theta)$\;
\end{algorithm}
\vspace{-0ex}\subsection{Generating Result Set via Learned Function}\label{ssec:generes}
The training process results in the learned parameters $\Theta$. Once the parameters are learned, they can be used to address \textsc{im} problem in multiple networks. In this part, we shall show the test phase of \textsc{disco} model, \ie online selection of seeds. First of all, as we have learned all the parameters in $\tilde{\sigma}(v,S;\Theta)$, we are able to directly calculate the predicted expected marginal influence for each node. Afterwards, intuitively, it is natural for one to follow the existing hill-climb strategy to iteratively select the best node that exhibits the highest predicted marginal influence. However, our empirical (shown in Section~\ref{ssec:exp2}) and theoretical studies (will show immediately) reveal that the order of the nodes with respect to their $Q$ values remain almost unchanged whenever we select a seed and recompute the network embeddings as well as the $Q$ values. Therefore, for the seeds selection phase, we hereby present a much better solution that is strictly suitable for our \textsc{disco} framework. To begin with, we shall first conduct a study over the intrinsic features for our learning model. In particular, with the help of Certainty-Factor (CF) model~\cite{books/aw/Jackson99}, we theoretically examine the difference of the predicted influences for any pair of nodes before and after some seed is selected into results set.
The CF model is a method for managing uncertainty in rule-based systems. A $\mathbf{CF}$ represents a person's change in belief in the hypothesis given the evidence. In particular, a $\mathbf{CF}$ from 0 to 1 means that the person's belief increases.

For an arbitrary pair of non-seed nodes $v_a$ and $v_b$, and current seeds set $S_i$, there corresponding predicted influences are $\tilde{\sigma}(v_a,S_i;\Theta)$ and $\tilde{\sigma}(v_b,S_i;\Theta)$, respectively. Suppose after a new seed, but not $v_a$ or $v_b$, is selected into the result set, leading to a new partial solution $S_{i+1}$. For brevity, we denote $\tilde{\sigma}(v_a,S_i;\Theta)$ and $\tilde{\sigma}(v_a,S_{i+1};\Theta)$ by $Q_A$ and $Q_A^\prime$, respectively. Similarly, we also use $Q_B$ and $Q_B^\prime$ for the values of node $v_b$. Then the following holds.

%

\begin{theorem}\vspace{-0ex}\label{lm1}
{\em $\forall v_a, v_b\in V$, where $v_a,v_b$ are not seeds, and a very small positive number $\eta$,
\begin{equation}\small
\begin{split}
\textbf{if  } &|(Q_A-Q_B)-(Q_A'-Q_B')|<\eta \\
\textbf{then  } &(Q_A-Q_B)(Q_A'-Q_B')>0, \textbf{CF} = 1
\end{split}
\end{equation}
That is, the order of any two nodes before and after recomputing the embeddings and $Q$ values does not change (The proof is given in Appendix~\ref{app:p1}).\/}
\vspace{-0ex}\end{theorem}

Further, we justify that $\eta$ is expected as a very small positive value.
\begin{claim}\vspace{-0ex}\label{lm2}
{\em $\forall v_a, v_b\in V$ and $v_a,v_b$ are not seeds, suppose the number of nodes and average degree of graph $G$ is $n$ and $|\overline{N(v)}|$, respectively. Then, according to \textsc{disco} model (where $I$ is fixed less than 5 and $Q$ values are Max-min Normalized):
\begin{equation*}\small
E[|(Q_A-Q_B)-(Q'_A-Q'_B)|]	<\frac{\sum_{i=1}^{4}|\overline{N(v)}|^i}{n}
\end{equation*} (The proof is given in Appendix~\ref{app:p2}).
\/}
\vspace{-0ex}\end{claim}
In practice, the average degree of nodes in real-world social networks are always very small (\ie within 3$\sim$7)~\cite{kumar2010structure} compared to the size of the network. Therefore, $\eta$ is expected to be a very small positive value in practice.

According to the above study, it is known that $|(Q_A-Q_B)-(Q_A'-Q_B')|$ should probably be a very small non-negative value. That is to say, whether or not we recompute the embeddings and $Q$ values after each seeds insertion can hardly affect the order of non-seed nodes. Therefore, during the seeds selection phase, instead of iteratively select-and-recompute the embeddings and $Q$ values according to each seed insertion, we simplify the procedure into only one iteration, by embedding only once and selecting the top-$k$ nodes with the maximum $Q$ without updating the predicted values.

The seeds selection process is outlined in Algorithm~\ref{alg:test}. Given a network $G$ and a budget $k$, we first initialize the seed set $S$ to be empty and initialize each node in the network as a $p$-dimensional zero vector (Lines 1). Next, we embed each node in the network (Lines 3-5). The update of the embeddings can be performed according to Equation~\ref{eq1}, and the parameters $\alpha_1,\ldots,\alpha_4$ have already been learned. The formula aggregates the neighborhood information of node $v$, and encodes the node information and network characteristics (Lines 5). After getting embeddings of the nodes, we calculate the influence quality of each node $v$ according to Equation~\ref{eq2} using the learned parameters $\beta_1,\ldots,\beta_3$ (Line 6). Finally, the node with the top-$k$ influence quality is added to the seed set.


The time complexity of the selection process is determined by three parts. First, in the embedding phase, the complexity is influenced by the number of nodes $\lvert V \rvert $ and the number of neighbors $\lvert \overline{N(v)} \rvert$. As $I$ is usually a small constant (\eg $I=4$) in Algorithm~\ref{alg:test}, network embedding takes $O(\lvert V\rvert \times \lvert \overline{N(v)}\rvert)$. Second, after the node is embedded, the quality of nodes in the graph is evaluated using the formula $Q(v,S,\Theta )= {\beta _1}^TReLU ([\beta _2\sum _{u\in V}x^{I} _u ,\beta _3x^{I} _v])$. Since the values of $\Theta $ have been learned, this step is influenced by the time taken to find the neighbors of node $v$. Hence, it takes $O(\lvert V\rvert \times \lvert\overline{N(v)}\rvert)$ time. Finally, selecting the optimal node according to $Q$ function and adding the node to the set $S$ takes $O(\lvert V\rvert)$. Consequently, the total time complexity for seeds selection is $O(\lvert V\rvert \times \lvert \overline{N(v)}\rvert)$.

\vspace{-0ex}\subsection{Pre-training the Model in Large-scale or Evolutionary Networks}\label{ssec:pretrain}
As a learning-based framework (also shown in Definition~\ref{pro:str2}), we need to pretrain the embedding and $Q$ function parameters offline using a group of training networks. Intuitively, the training and testing (\ie target) networks should be homogeneous in terms of topology such that the quality of the learned parameters can be guaranteed. In general, the homogeneity in terms of topology can be reflected in the aspects of many topological properties, \eg degree distribution, spectrum, etc. Therefore, in order to select seeds within a targeted network, we need to train the required parameters $\Theta$ in a group of homogeneous networks offline. Afterwards, the trained model can be further used to select seeds in a target network. In the following, depending on the specific characteristic of target networks, we shall present two different pretraining strategies of \textsc{disco}.

\textbf{Applying \textsc{disco} in large-scale stationary networks.}
Consider a large-scale stationary network without any evolution log. We can turn to subgraph-sampling technique to generate sufficient homogeneous training networks. Afterwards, with learned parameters $\Theta$ from these small networks, we can address \textsc{im} over the target large-scale network. In order to ensure that the topological features of the sampled training subgraphs are as consistent as possible with the original large-scale target network, we evaluate different sampling algorithms following the same framework adopted in \cite{DBLP:journals/tkdd/AhmedNK13,DBLP:conf/kdd/LeskovecF06}. In particular, we apply different sampling methods to real large-scale networks and sample subgraphs over a series of sample fractions $\varphi$ = $[1\%, 5\%, 10\%, 15\%]$. Following the methods introduced in~\cite{DBLP:journals/tkdd/AhmedNK13}, Kolmogorov-Smirnov D-statistic is adopted to measure the differences between the network statistics distributions of the sampled subgraphs and the original graph, where D-statistic is typically applied as part of the Komogorov-Smirnov test to reject the null hypothesis. It is defined as $D = \max_x \{| F_{0}(x) - F(x)|\}$, where $x$ denotes the range of random variables; $F_{0}$ and $F$ are two empirical distribution functions of the data. We found that our experimental results show similar phenomenon with the benchmarking papers \cite{DBLP:journals/tkdd/AhmedNK13,DBLP:conf/kdd/LeskovecF06}. That is, Topology-Based Sampling is superior to Node Sampling and Edge Sampling in the distribution of graph features such as degree distribution and clustering coefficient distribution. So in the next section, we adopt Topology-Based sampling methods in our model and compare several existing Topology-Based subgraph sampling methods, including Breadth First Sampling (BFS)\cite{Doerr:2013gz}, Simple Random Walk (SRW) and its variants, namely Random Walk with Flyback (RWF) and Induced Subgraph Random Walk Sampling (ISRW) \cite{Lee:2012ga,Leskovec:2006fk}, as well as Snowball Sampling (SB)\cite{Goodman:1961fh}, a variant of Breadth First Search which limits the number of neighbors that are added to the sample. Notably, as the subgraph-sampling technique is beyond the focus of this work, we do not discuss the detailed techniques for these methods.

\textbf{Applying \textsc{disco} in evolutionary networks.} Almost all existing social networks keep on evolving, with insertion of new nodes/edges and deletion of old nodes/edges. Although social networks evolve rapidly over time, the structural properties remain relatively stable~\cite{kumar2010structure}. During the evolution of a particular real-world network, we advocate that any two historical snapshots of the same network are homogeneous to each other. By leveraging on the aforementioned technique, we now briefly describe how \textsc{disco} can address the \textsc{im} problem in dynamic networks via time-based sampling.
In particular, given a dynamic network $G$, whose snapshot at time $t$ is referred to as  $G_t$, we can apply \textsc{disco} in the following way. During the \textit{Learning phase} of Definition~\ref{pro:str2}, we can train the model using a series of temporal (sampled) network snapshots (\ie $\mathcal{G}=\{G_{t1},\ldots,G_{t\ell}\}$). For the \textit{Seeds selection phase}, the trained model can be used to select the best seed set in an arbitrary snapshot of $G$. Besides, if any snapshot of the network is very large, we can also adopt the subgraph sampling method mentioned above.

Based on this strategy, we can accomplish the seeds selection task over a target network in real-time, although it keeps on evolving. This provides a practical solution for the \textsc{im} problem in evolutionary networks.
\begin{figure*}[t]
	\vspace{0ex}\centering
	\subfloat[][$\textbf{HepPh}$ \label{f4a}]{\includegraphics[clip=true, width=0.25\linewidth]{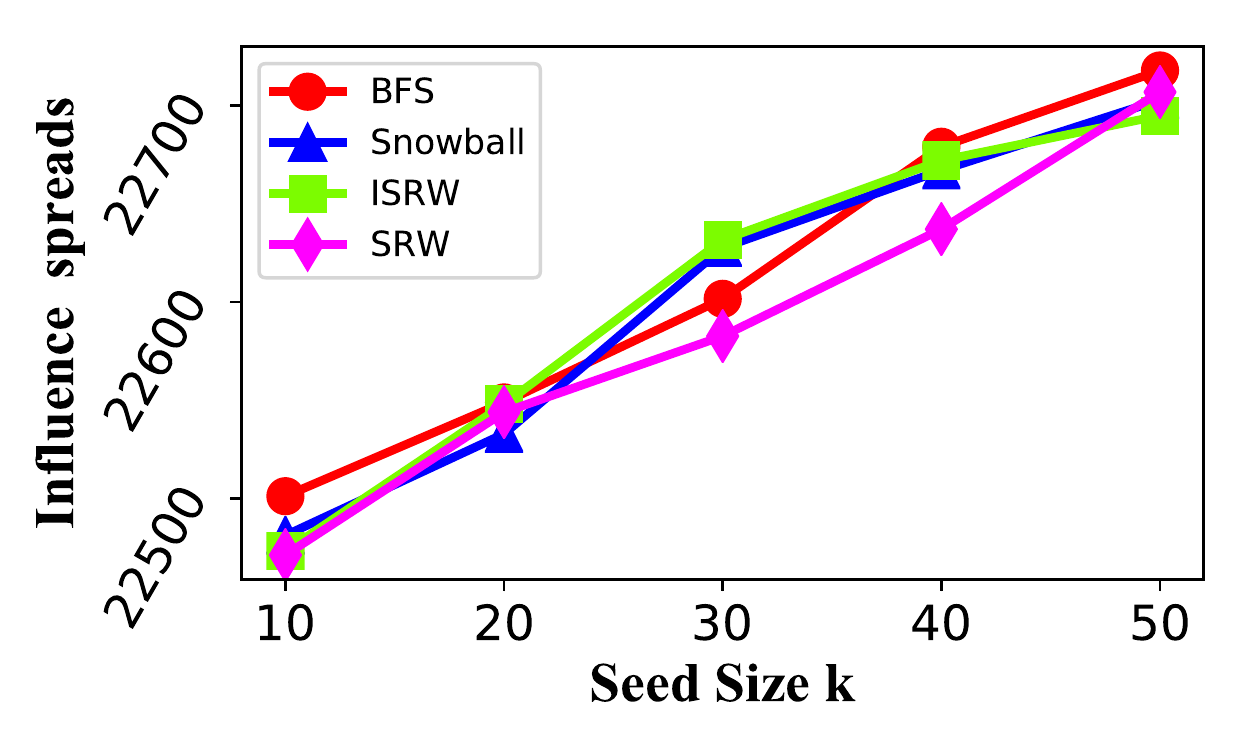}}
	\subfloat[][$\textbf{DBLP}$ \label{f4b}]{\includegraphics[clip=true, width=0.25\linewidth]{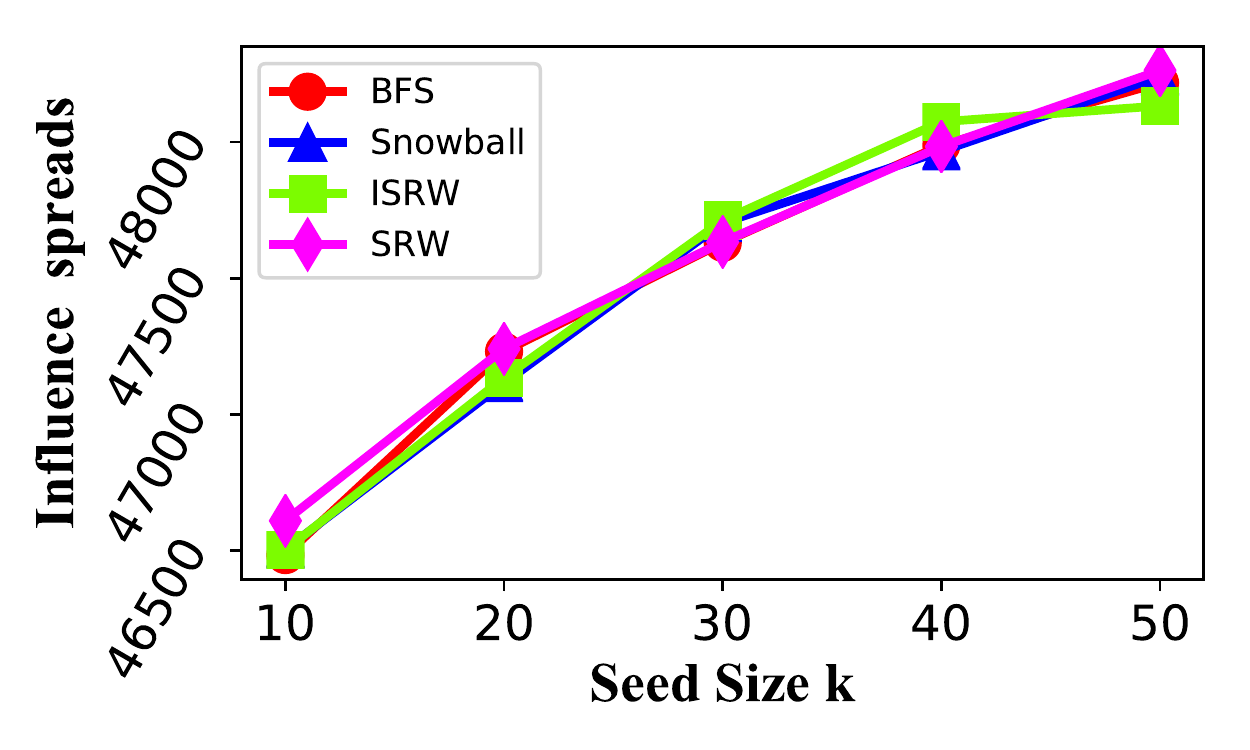}}
	\subfloat[][$\textbf{LiveJournal}$ \label{f4c}]{\includegraphics[clip=true, width=0.25\linewidth]{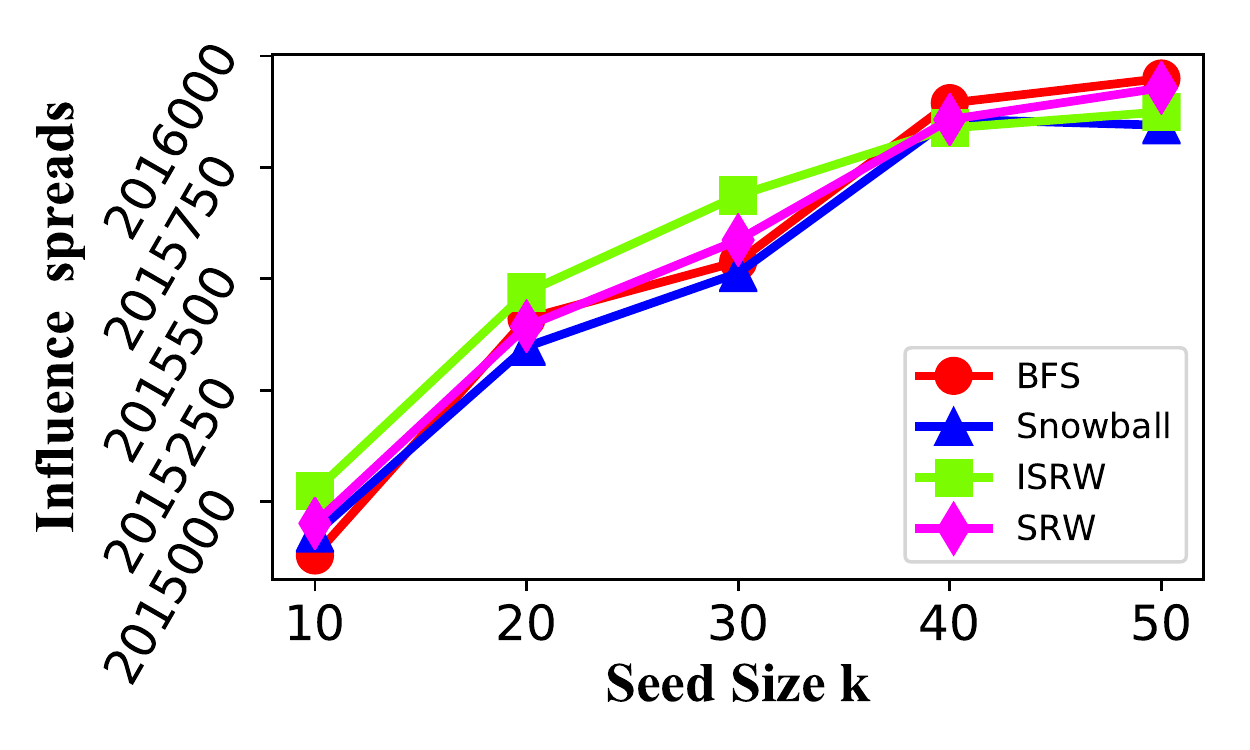}}
	\subfloat[][$\textbf{Orkut}$ \label{f4d}]{\includegraphics[clip=true, width=0.25\linewidth]{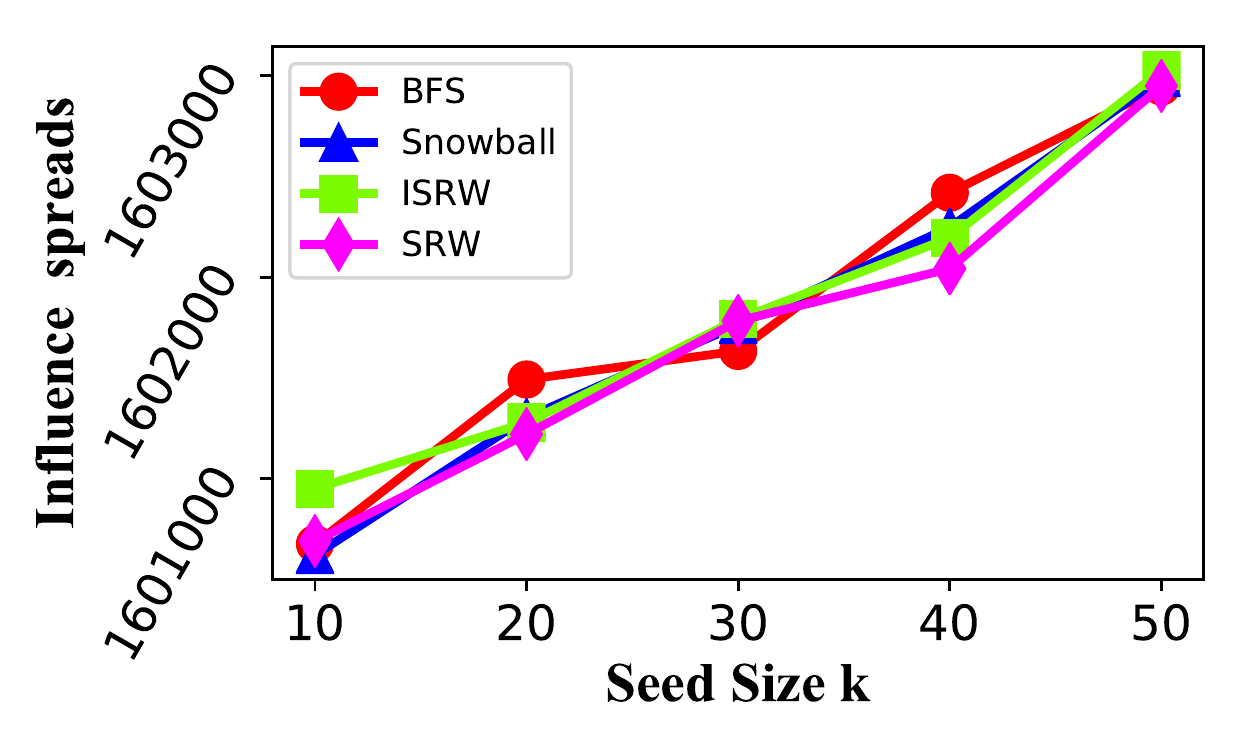}}
	\vspace{-0ex}\caption{Influence spreads with different sampling algorithms.}\label{sampling}
	\vspace{-0ex}\vspace{-1ex}
\end{figure*}

\vspace{0ex}\section{Experiments}\label{sec:exp}

In this section, we evaluate the performance of the \textsc{disco} framework. We compare our model with two state-of-the-art traditional \textsc{im} solutions, namely \textsc{imm} \cite{Tang2015} and \textsc{ssa} \cite{DBLP:journals/corr/NguyenTD16}, as suggested by a recent benchmarking study~\cite{DBLP:conf/sigmod/AroraGR17}. Recall that the efforts in~\cite{DBLP:conf/webi/AliWC18,DBLP:conf/kdd/LinLC15,DBLP:conf/kdd/SunHYC18,DBLP:journals/pvldb/HanHXTST18} are designed for competitive \textsc{im} and hence are orthogonal to our problem. In line with all the \textsc{im} solutions, we evaluate the performances from two aspects, namely computational efficiency and influence quality. Besides, as a learning-based solution, we also justify our model generality within evolutionary scenarios. All the experiments were performed on a machine with Intel Xeon CPU (2.2GHz, 20 cores), 512GB of DDR4 RAM, Nvidia Tesla K80 with 24GB GDDR5, running Linux\footnote{We shall provide the url of complete source code for \textsc{disco} here as soon as this paper is accepted.}.

\begin{table}[!t]
\centering
\caption{Datasets}\label{tab:dataset}
\begin{tabular}{ccccc}
\toprule
\textbf{Dataset}& \textbf{n}& \textbf{m}&\textbf{Type}& \textbf{Avg.Degree}\\
\midrule
\textit{HepPh} & 12K &118K &Undirected &9.83\\
\textit{DBLP} & 317K &1.05M &Undirected &3.31\\
\textit{LiveJournal} & 4.85M &69M &Directed &6.5\\
\textit{Orkut} & 3.07M &117.1M &Undirected &4.8\\
\bottomrule
\end{tabular}
\vspace{-1ex}\end{table}

\begin{figure*}[t]
	\vspace{0ex}\centering
	\subfloat[][$\textbf{HepPh}$ \label{f4a}]{\includegraphics[clip=true, width=0.25\linewidth]{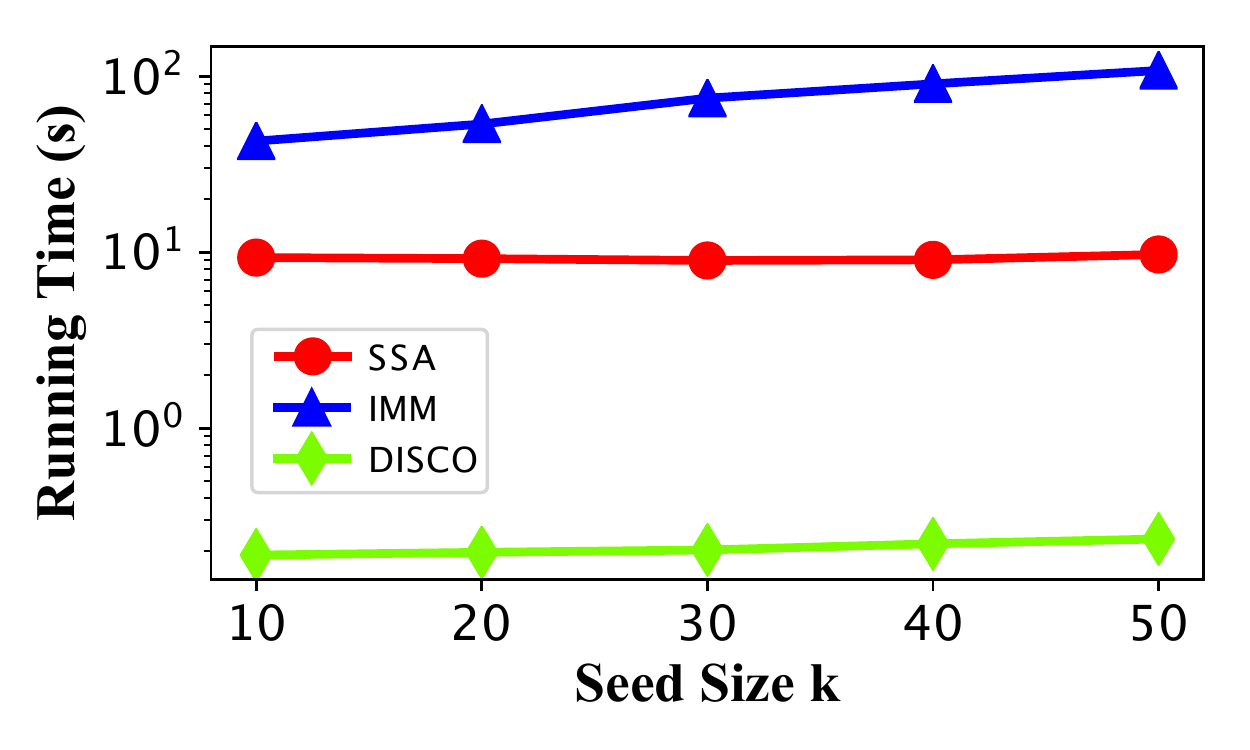}}
	\subfloat[][$\textbf{DBLP}$ \label{f4b}]{\includegraphics[clip=true, width=0.25\linewidth]{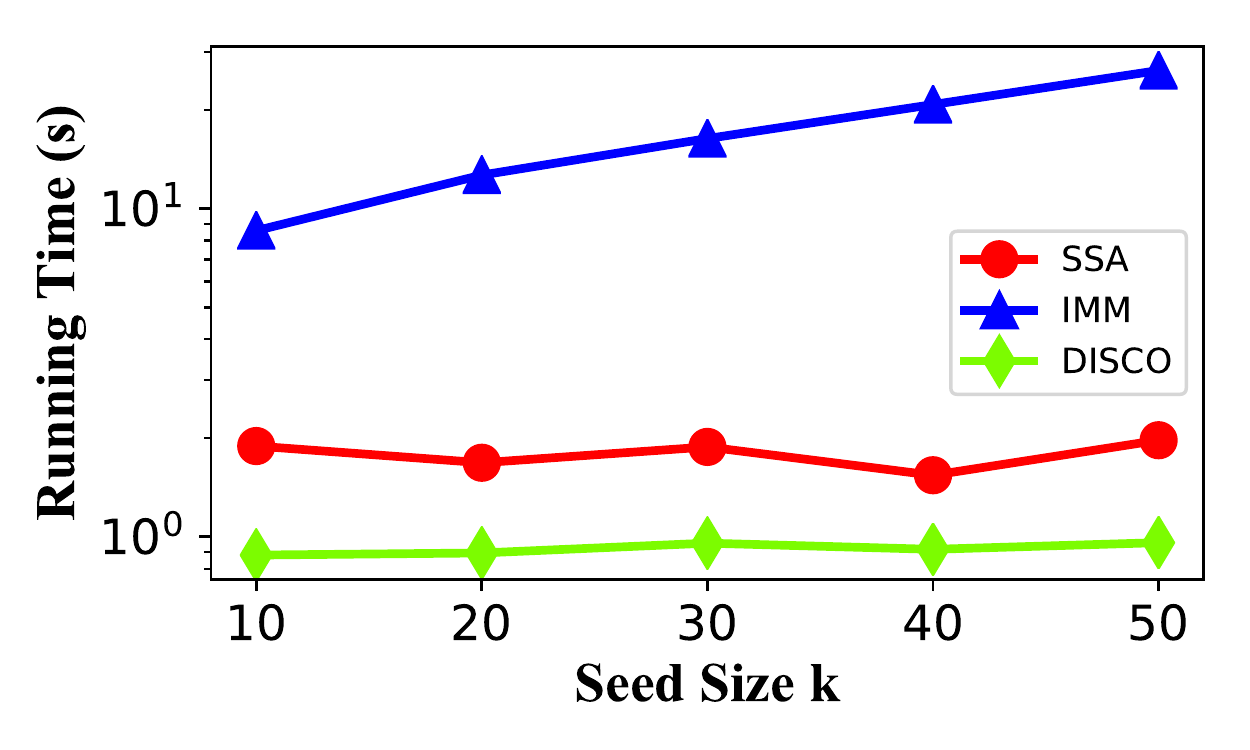}}
	\subfloat[][$\textbf{LiveJournal}$ \label{f4c}]{\includegraphics[clip=true, width=0.25\linewidth]{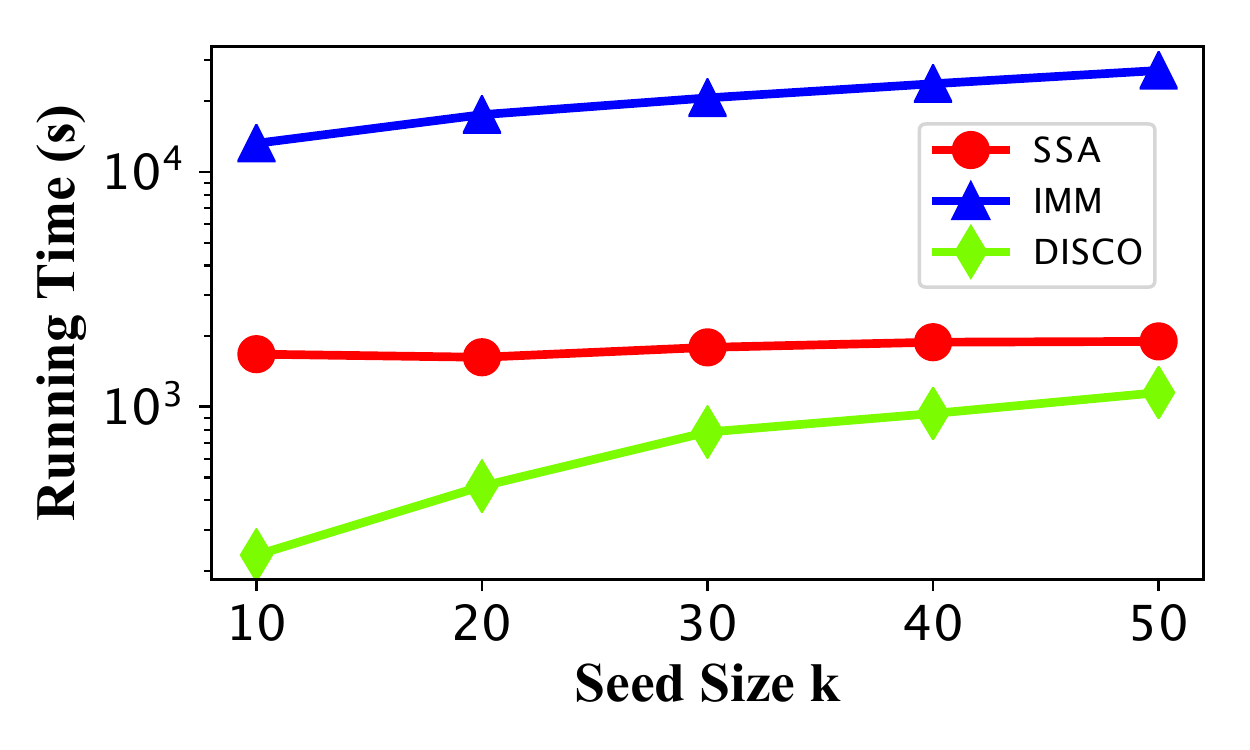}}
	\subfloat[][$\textbf{Orkut}$ \label{f4d}]{\includegraphics[clip=true, width=0.25\linewidth]{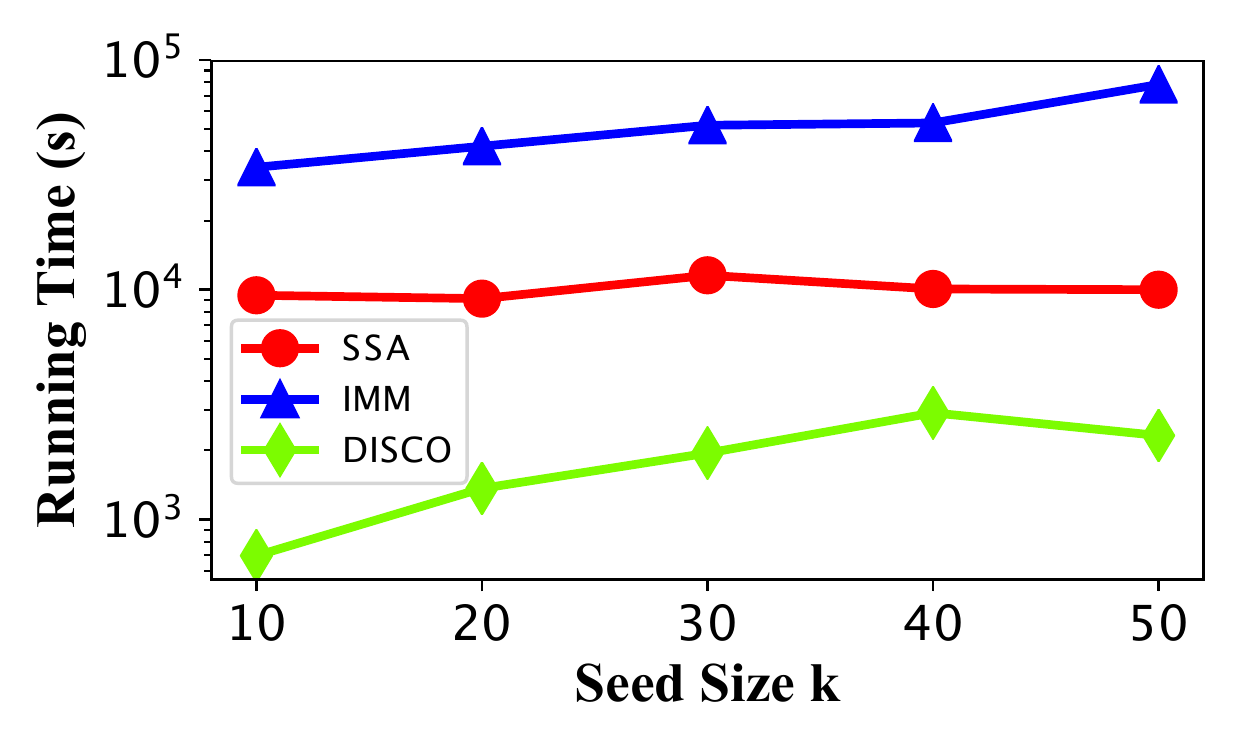}}
	\vspace{-0ex}\caption{Running time.}\label{time}
	\vspace{-0ex}\vspace{-1ex}\end{figure*}
\begin{figure*}[t]
	\vspace{0ex}\centering
	\subfloat[][$\textbf{HepPh}$ \label{f4a}]{\includegraphics[clip=true, width=0.25\linewidth]{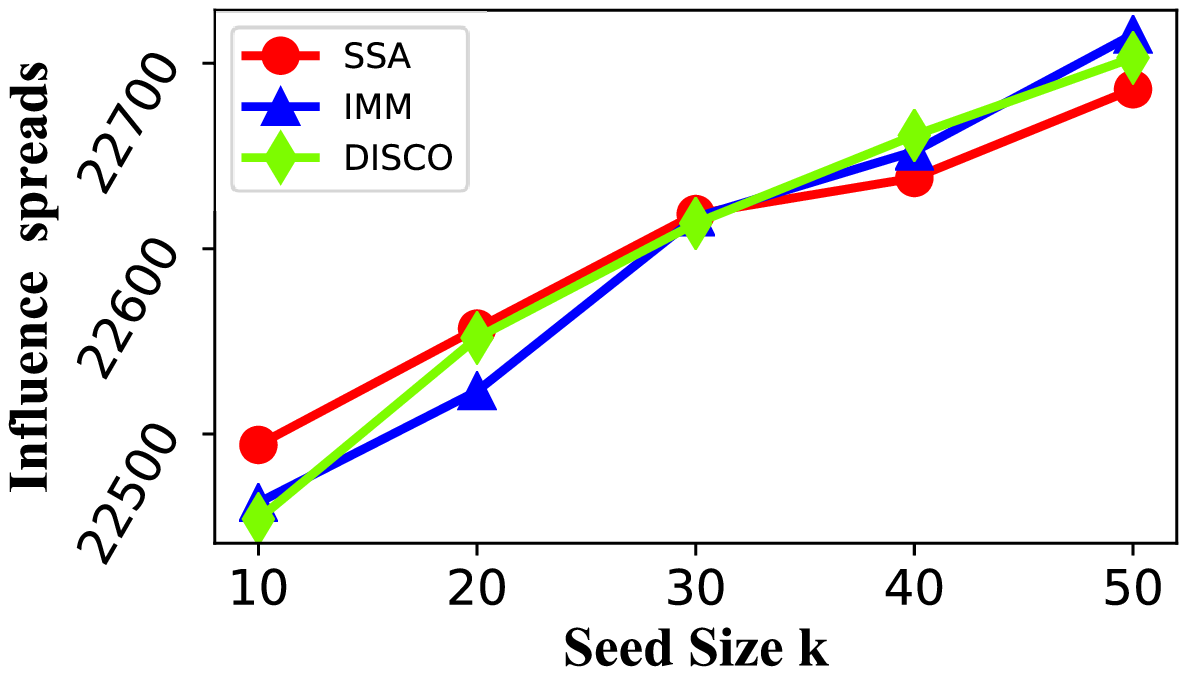}}
	\subfloat[][$\textbf{DBLP}$ \label{f4b}]{\includegraphics[clip=true, width=0.25\linewidth]{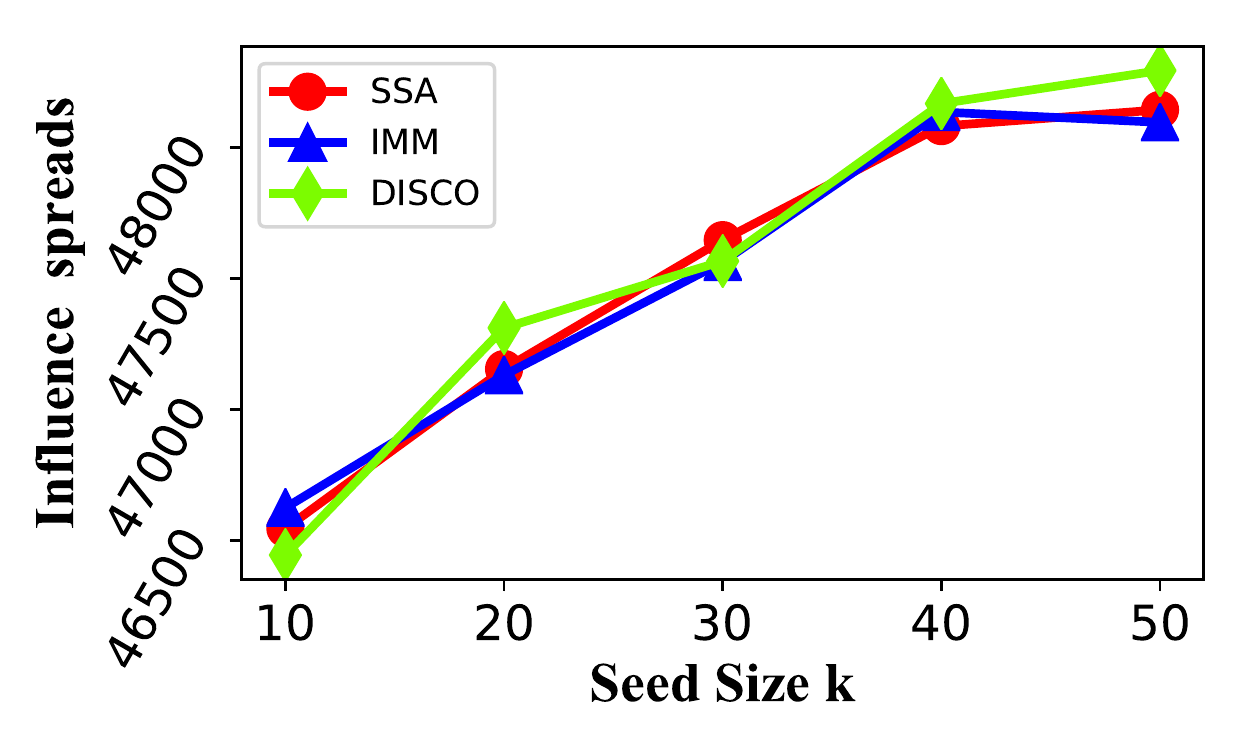}}
	\subfloat[][$\textbf{LiveJournal}$ \label{f4c}]{\includegraphics[clip=true, width=0.25\linewidth]{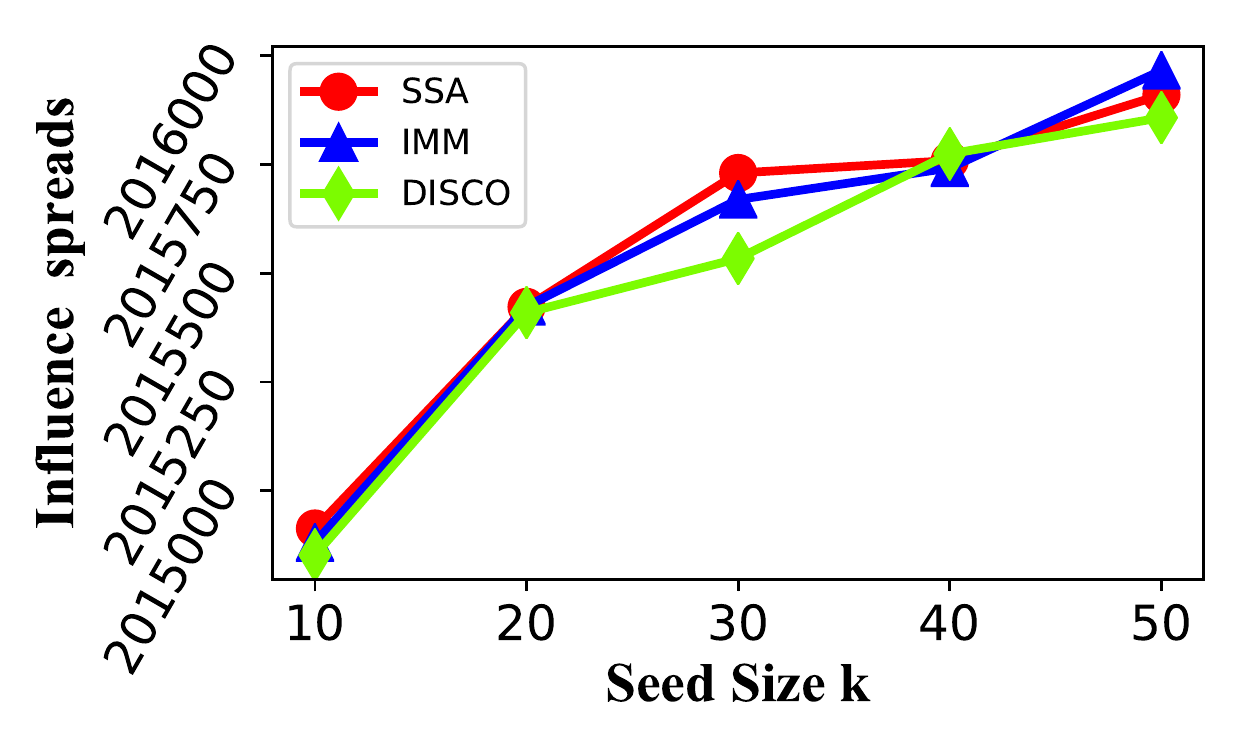}}
	\subfloat[][$\textbf{Orkut}$ \label{f4d}]{\includegraphics[clip=true, width=0.25\linewidth]{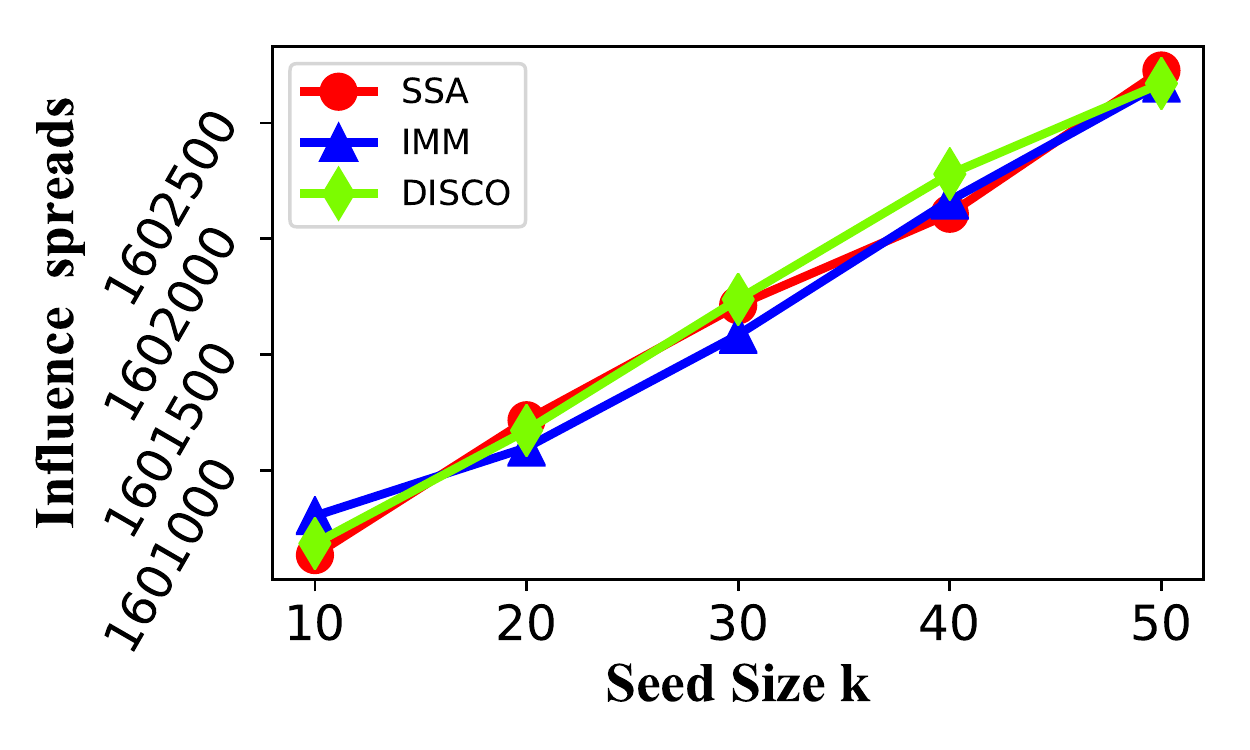}}
	\vspace{-0ex}\caption{Influence spreads quality.}\label{inf}\vspace{-0ex}\vspace{-1ex}
\end{figure*}

\vspace{0ex}\subsection{Experimental Setup}
\textbf{Datasets.} In the experiments, we present results on four real-world social networks, taken from the \textit{SNAP} repository\footnote{\url{https://snap.stanford.edu/data/}}, as shown in Table \ref{tab:dataset}. In these four datasets, \textit{HepPh} and \textit{DBLP} are citation networks, \textit{LiveJournal} and \textit{Orkut} are the largest online social networks ever used in influence maximization. For each real-world network, we use the following sampling methods: Breadth First Sampling (BFS)~\cite{Doerr:2013gz}, Simple Random Walk (SRW), Induced Subgraph Random Walk Sampling (ISRW) \cite{Lee:2012ga,Leskovec:2006fk}, and Snowball Sampling (SB)~\cite{Goodman:1961fh}.

\textbf{Diffusion Models.} \textsc{disco} can be easily adapted to different diffusion models. For instance, we can simply revise the Reward definition to switch from \textsc{ic} model to \textsc{lt} model. As our experimental results under both \textsc{lt} and \textsc{ic} models are qualitatively similar, we mainly report the results under the \textsc{ic} model here. In \textsc{ic} model, each edge $W(u, v)$ has a constant probability $p$. In vast majority of the \textsc{im} techniques, $W(u, v)$ takes the value of $0.5$ assigned to all the edges of the network. In order to fairly calculate the expected range of influence for the three approaches, we first record the seed set of each algorithm independently, and then perform $10,000$ simulated propagations based on the selected seeds. Finally, we take the average result of $10,000$ simulations as the number of influenced nodes for each tested approach.

\textbf{Parameter Settings.} For our method, we set the batch size, \ie the number of samples extracted from $M$ each time, as $64$. Besides, we set $\delta$ as 5 and the learning rate of \textsc{sgd} to $0.001$. We set the dimension of the nodes embedded in the network to $64$. As suggested in \textsc{imm} \cite{Tang2015}, we set $\epsilon= 0.1$ throughout the experiments in both \textsc{imm} and \textsc{ssa}. It should be noted that the seed set produced by \textsc{ssa} is not constant. Therefore, for \textsc{ssa} and \textsc{imm}, we report the average results over 100 independent runs (\ie 100 independent results seed sets, each of which is averaged for $10,000$ simulations).

\vspace{-0ex}\subsection{Experimental Results}\label{ssec:exp2}
\textbf{Training issues.}
As remarked earlier, during \textsc{dqn} training, we need to obtain a batch of training graphs by sampling the real-world network. In our experiments, we mainly compare the aforementioned four sampling methods (\textsc{BFS}, \textsc{ISRW}, Snowball, and \textsc{SRW}). The results are shown in Figure \ref{sampling}. Observe that for the \textsc{disco} training process, the results of the subgraph training model obtained by using different Topology-Based sampling methods have ignorable difference, indicating that different Topology-Based sampling method has little effect on the result seeds quality. In other words, the Topology-Based sampling method and the \textsc{disco} framework are loosely coupled. Therefore, we chose to use simple and mature \textsc{BFS} (Breadth First Sampling) as the default method in rest of the experiments.

Given the sampled sub-networks, we train the \textsc{dqn} parameters using $\varepsilon$-greedy exploration described in Section~\ref{ssec:modeltrain}. That is, the next action is randomly selected with probability $\varepsilon$, and the action of maximizing the $Q$ function is selected with probability $1-\varepsilon$. During training, $\varepsilon$ is linearly annealed from 1 to 0.05 in ten thousand steps. The training procedure will terminate when the value of $\varepsilon$ is less than $0.05$ or the training time exceeds 3 days. For both \textit{HepPh}, \textit{DBLP}, the training phases terminate at 2.5 and 23.5 hours, respectively; for \textit{LiveJournal} and \textit{Orkut}, we limit the training time within 3 days and apply the learned parameters to node selection.  

Notably, \textit{the training phase is performed only once and we do not need to train while running an \textsc{im} query}. The study on evolutionary networks at the end of this section further justifies that, once trained, it can be applied many times to address the \textsc{im} problem.

\textbf{Computational efficiency.}
We compare the running times of the three algorithms by varying $k$ from $10$ to $50$. The results on datasets \textit{HepPh}, \textit{DBLP}, \textit{LiveJournal} and \textit{Orkut} are reported in Figure~\ref{time}. We can make the following observations. In terms of computational efficiency, \textsc{disco} significantly outperforms \textsc{imm} and \textsc{ssa} by a huge margin. Regardless of the value of $k$, the cost of the \textsc{disco} is significantly less than that of \textsc{imm} and \textsc{ssa}. Particularly, when the number of seed nodes selected is small, the performance gain is more prominent. For example, under the \textsc{ic} model, when $k = 10$ and $k=50$, \textsc{disco} is 23 and 1.57 times faster than \textsc{ssa} on \textit{Live-Journal} network, respectively.

\textbf{Influence quality.}
Next, we compare the quality of the seed sets generated by \textsc{imm}, \textsc{ssa} and \textsc{disco} on \textsc{IC} models. For each algorithm we set the optimal parameter values according to their original papers and then evaluate their quality. As can be seen from Figure \ref{inf}, our proposed method is as good as \textsc{imm} and \textsc{ssa} in real networks of different sizes, and the growth of influence spread with $k$ have few minor fluctuations. Note that the scales of the vertical axis in the four figures are different. In our experiments, we also observe that \textsc{ssa} results are unstable. In comparison, our method can choose superior result set, and the results are stable across $100$ runs. Importantly, \textsc{disco} produces the same or even better quality results as the state-of-the-art.

\begin{table*}[ht]
	\small
	\centering
	\caption{Effectiveness of our seeds selection strategy.}\label{seed_order}\vspace{-1ex}
	\begin{tabular}{|c|c|c|c|c|c|c|c|c|}
		
		\hline
		&\multicolumn{2}{|c|}{$k=1$} & \multicolumn{2}{|c|}{$k=10$} & \multicolumn{2}{|c|}{$k=25$} & \multicolumn{2}{|c|}{$k=50$}\\
		\hline
		& $\Delta rank$ & $\Delta Inf$ & $\Delta rank$ & $\Delta Inf$ & $\Delta rank$ & $\Delta Inf$ & $\Delta rank$ & $\Delta Inf$\\
		
		\hline
		\textit{HepPh} & 1 & 1 & 0.997 & 0.996 & 0.970 & 0.996 & 0.95 & 0.995 \\
		\textit{DBLP} & 1  &1 & 0.942 & 0.998 & 0.962 & 0.998 & 0.968 & 0.996  \\
		\textit{LiveJournal} & 1 & 1 & 1 & 0.999 & 0.985 & 0.998 & 0.979 & 0.998 \\
		\textit{Orkut} & 1 & 1 & 0.983 & 0.995 & 0. 965 & 0.997 & 0.972 & 0.993 \\
		\hline
	\end{tabular}
\end{table*}	

\textbf{Justification of seeds selection strategy.}
When we select the seed set, we only execute the process of network embedding and then select $k$ nodes with the largest $Q$ values, instead of updating the embedding every time a seed node is selected. The purpose of this experiment is to show the difference over the results between the strategy shown in Section~\ref{ssec:generes} and traditional hill-climb strategy, \ie select-and-update marginal influence iteratively. To this end, we compare the results by these two different strategies and report the differences in terms of node order and influence quality. We hereby report the results in Table~\ref{seed_order}. The column $\Delta Inf$ records the difference between the result quality for our seeds selection strategy of Section~\ref{ssec:generes} and the traditional hill-climb iterative method (\ie select a seed and update the marginal influence for the rest iteratively). The column titled $\Delta rank$ records the probability that the node order changes between the two strategies. As can be seen from Table \ref{seed_order}, the probability of the order does not change in a seed set when we select one node iteratively or select $50$ nodes at a time ($0.972$ on \textit{Orkut}). Once again, this supports our theoretical discussion in Theorem~\ref{lm1} and Claim~\ref{lm2}.

\vspace{-0ex}\begin{figure}[t]
	\vspace{-0ex}\centering
	\includegraphics[clip=true, width=0.65\linewidth]{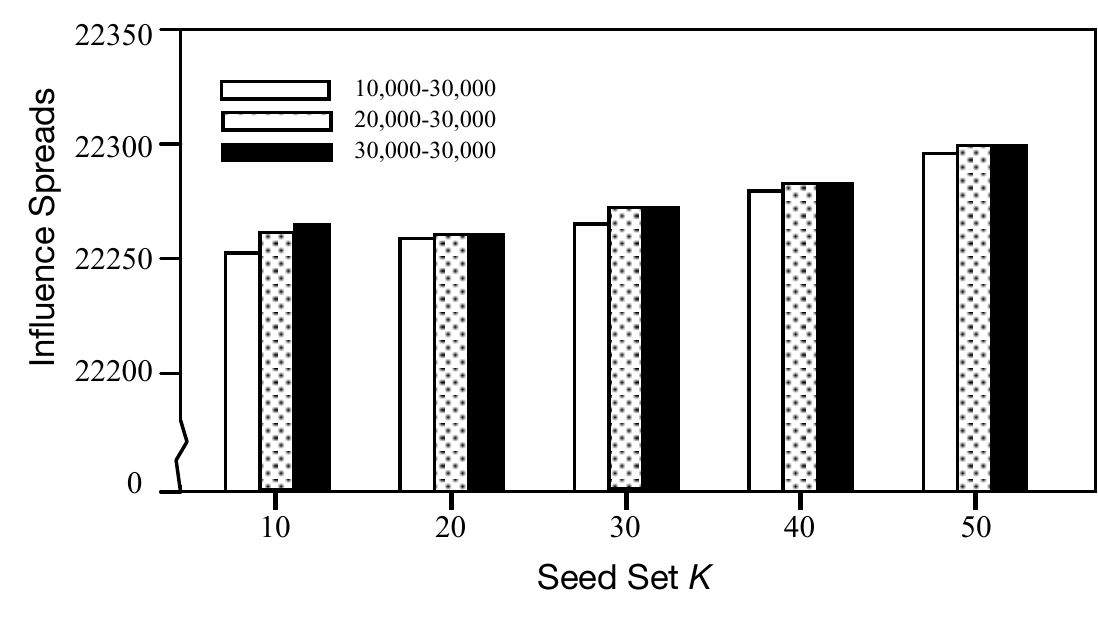}
	\vspace{-1ex}\caption{Performance in real-world evolutionary networks.}\label{fig:fit}\vspace{-1ex}
\end{figure}
\textbf{Performance in evolutionary networks.} Finally, we evaluate the performance of \textsc{disco} on evolutionary networks following the strategy discussed in Section~\ref{ssec:pretrain}. Among all the tested datasets, \textit{HepPh} is associated with evolution logs, \ie the time-stamps when each node/edge is inserted to the network. Therefore, following the proposed implementation model of \textsc{disco} in evolutionary network, we train the model using a series of temporal snapshots for \textit{HepPh} networks when it has $10,000, 20,000$, and $30,000$ nodes, respectively. Then the three trained models are tested on a future snapshot of the network, which contains $30,000$ nodes, during the evolution. As can be seen in Figure \ref{fig:fit}, the trained model from $10,000$-nodes snapshot and that from $30,000$-nodes ones have little differences in terms of the influence quality. For instance, when the number of seeds is $50$, the difference between $10,000$-nodes model and $30,000$-nodes model in the influence spread is $224$, which is about $1\%$ of the total influenced nodes. Therefore, in real-world dynamic networks, \textsc{disco} can be easily trained using earlier snapshot or even the subnetworks of some snapshots. When the network evolves and expands, we can continue to use the pretrained model for selecting seeds in a larger network snapshot.

\vspace{0ex}\section{Conclusions}\label{sec:concl}
In this work, we have presented a novel framework, \textsc{disco}, to address the \textsc{im} problem by exploiting machine learning technique. To the best of our knowledge, we are the first to employ deep learning methods to address the classical \textsc{im} problem. Our framework incorporates both \textsc{dqn} and network embedding methods to train superior approximation of the $\sigma$ function in \textsc{im}. Compared to state-of-the-art sampling-based \textsc{im} solutions, \textsc{disco} can avoid the costly diffusion sampling phase. By training with sampled subnetworks once, the learned model can be applied many times. Therefore, we are able to achieve superior efficiency compared to state-of-the-art classical solutions. Besides, the result quality in terms of influence spread of \textsc{disco} is the same or better than the competitors. As the seminal effort towards the paradigm of learning-based \textsc{im} solution, \textsc{disco} demonstrates exciting performance in terms of result quality and running time. We believe that our effort paves the way for a new direction to address the challenging classical \textsc{im} problem.
\section*{Acknowledgment}\label{sec:ack}
The work is supported by National Nature Science Foundation of China (No. 61672408), the Director Foundation Project of National Engineering Laboratory for Public Safety Risk Perception and Control by Big Data (PSRPC), Fundamental Research Funds for the Central Universities (No. JB181505), Natural Science Basic Research Plan in Shaanxi Province of China (No. 2018JM6073) and China 111 Project (No. B16037).
\appendices
\section{Proof of Theorem \ref{lm1}}\label{app:p1}	
Given $|(Q_A-Q_B)-(Q_A'-Q_B')|<\eta$, there are two different cases as follows.

\textbf{Case 1}: $(Q_A-Q_B)-(Q_A'-Q_B')\ge 0$, then it is $(Q_A-Q_B)-(Q_A'-Q_B')<\eta$;

\textbf{Case 2}: $(Q_A-Q_B)-(Q_A'-Q_B')< 0$, then it is $(Q_A-Q_B)-(Q_A'-Q_B')>-\eta$.

We shall prove the theorem in \textit{Case 1 }first.

First we assume that at state $S$, given $Q$ value of every node, which are ranked in the following order:
	\begin{multline*}
    Q_1 > Q_2 > \cdots >Q_c>\cdots>Q_i>\cdots > Q_j > \cdots > Q_n
	\end{multline*}
For any nodes $i$ and $j$, $Q_i-Q_j>\eta$, where $\eta$ is a small positive value. We find that after selecting a seed node $c$, the $Q$ values recalculated after embedding are updated to the following values, respectively:
	\begin{equation*}
		Q'_1,Q'_2,\cdots,Q'_i,\cdots,Q'_j,\cdots,Q'_{(n-1)}
	\end{equation*}
As $(Q_i-Q_j )-(Q'_i-Q'_j)<\eta$, we calculate the probability for $Q_i'$ and $Q_j'$ to keep their order, that is to calculate $P(Q'_i-Q'_j>0)$. Besides, $(Q_i-Q_j )-(Q'_i-Q'_j)<\eta$ is equivalent to $(Q'_i-Q'_j)>(Q_i-Q_j )-\eta$. To get $P(Q'_i-Q'_j>0)$, the above probability can be inferred by scaling to
\begin{equation*}
	P(Q'_i-Q'_j>0) > P((Q_i-Q_j)>\eta)
\end{equation*}
As $\eta>0$, $P((Q_i-Q_j)>\eta)$, by the condition $Q_i-Q_j>\eta$, $P((Q_i-Q_j)>\eta)=1$ can be obtained, then $P(Q'_i-Q'_j>0) =1$.

For \textit{Case 2}, the proof is the same as\textit{ Case 1}.
\section{Proof of Claim \ref{lm2}}\label{app:p2}
The proof can be separated into two stages.	

\noindent\textbf{First stage.} Firstly, we can prove that when $I=4$,
\begin{equation}\label{eq:stage1}\small
|(Q_A-Q_B)-(Q'_A-Q'_B)|=|\beta''^T_1 \beta_3\alpha _2\alpha _3\sum_{i=1}^4\alpha _1^{i-1}[(\mathcal{N}_A^{(S')(i)}-\mathcal{N}_A^{(S)(i)})-(\mathcal{N}^{(S')(i)}_B-\mathcal{N}_B^{(S)(i)})]|
\end{equation}
where $\mathcal{N}_A^{(S')(i)}$ refers to the sum of the embeddings for nodes that are $i$-th hop away from node $A$ when all nodes in $S$ have been removed from the graph.

According to the update process of $Q$ at state $S$
	\begin{equation}\label{upQ}\small
	Q_v={\beta_1}^TReLU ([\beta _2\sum _{u\in V}x^{(S)(I)} _u,\beta _3x^{(S)(I)} _v])
	\end{equation}
 $\beta_1^T \in \mathbb{R}^{1\times 2p}$, and $\beta_2^T$,$\beta_3^T \in \mathbb{R}^{p\times p}$.
We divide $\beta_1^T$ into two parts, the former $p$ column is $\beta_1^{'T}$, and the last $p$ column is $\beta_1^{''T}$. Thus the update formula for $Q_v$ at state $S'$ is:
\begin{equation}\label{up1}\small
	Q'_v=\beta_1^{''T} \textit{ReLU}\left(\beta_2\sum_{u\in V}x_u^{(S')(I)}\right)+\beta_1^{''T}\textit{ReLU}(\beta_3\  x_v^{(S')(I)})
\end{equation}
We can get it from Eq.~\ref{up1} as follows.
\begin{equation*}\small
\begin{split}	P(|(Q_A-Q_B)-(Q'_A-Q'_B)|<\epsilon)=
&P(|[{\beta ''}^T_1(\textit{ReLU}({\beta}_3x_A^{(S)(I)})-\textit{ReLU}(\beta_3x_B^{(S)(I)}))]\\
&-[{\beta''}_1^T(\textit{ReLU}(\beta_3x_A^{(S')(I)})-\textit{ReLU}(\beta_3x_B^{(S')(I)}))]|)
\end{split}
\end{equation*}

Notably, in the initial state, all parameters are random numbers in the interval $(0, 0.1)$. Besides, all the parameters in $\Theta$ are vectors, whose entries are non-negative (in implementation they are all normalized into [0,1]). Without loss of generality, we assume all these parameters are $p$-dimensional. Then, $\textit{ReLU}$ can be resolved to get
\begin{equation}\label{eq15}\small
P(|(Q_A-Q_B)-(Q'_A-Q'_B)|<\eta)=
P\left(\left|{\beta ''}^T_1 \beta_3 \left[(x_A^{(S)(I)}-x_B^{(S)(I)})-(x_A^{(S')(I)}-x_B^{(S')(I)})\right]\right| < \eta \right)
\end{equation}
Hereby, $ x_v^{(S)(I)}$ represents the final vector after $I$ iterations for state $S$.
During the current state, $A$ and $B$ are not seed nodes, so the corresponding $a_v$ is $0$. Next, we will use $x_A^{(S')}$ as an example to expand the equation. $x_A^{(S)}, x_B^{(S)}, x_B^{(S')}$ are the same as $x_A^{(S')}$, and are not described here. Then the update formula for each iteration is
\begin{equation}\label{eq7}
\small
	x_A^{(S')(4)}=\alpha _1 \sum_{u\in N(A)} x_u^{(S')(3)} +\alpha _2 \alpha _3 \sum_{u \in N(A)} w(u, A)
\end{equation}

Assume that $A$ have $\kappa$ neighbors, which are $u_1,\cdots,u_\kappa$, then Eq. \ref{eq7} can be expanded to:
\begin{equation}
\small
	x_A^{(S')(4)}=(\alpha _1 \left(\mu_{u_1}^{(S')(3)} +\cdots+\mu_{u_\kappa}^{(S')(3)}\right)
	+\alpha _2 \alpha _3 \sum_{u \in N(A)} w(u, A)
\end{equation}

Similarly, we assume that $u_\kappa$ have $\lambda$ neighbors, namely $u_{\kappa(1)},\cdots,u_{\kappa(\lambda)}$. For $u_{\kappa(\lambda)}$, we assume that there are $\mu$ neighbors, namely $u_{\lambda(1)}, \cdots, u_{\lambda(\mu)}$. For $u_{\lambda(\mu)}$, we assume that there are $\nu$ neighbors, namely $u_{\mu(1)},\cdots, u_{\mu(\nu)}$. Because in our experiment, $x_v^{(S')(0)}=0$ of all nodes, so that can be expressed as follows:
\begin{equation*}
\small
\begin{split}
	x_A^{(S')(4)}=
	&\alpha _1^3\alpha _2\alpha _3 \sum_{u_\kappa \in N(A)}\sum_{u_{\kappa(\lambda)} \in N(u_\kappa)}\sum_{u_{\lambda(\mu)} \in N(u_{\kappa(\lambda)})}\sum_{u_{\mu(\nu)} \in N(u_{\lambda(\mu)})} w(u_{\mu(\nu)},u_{\lambda(\mu)})\\
&+\alpha _1^2\alpha _2\alpha _3 \sum_{u_\kappa \in N(A)}\sum_{u_{\kappa(\lambda)} \in N(u_\kappa)}\sum_{u_{\lambda(\mu)} \in N(u_{\kappa(\lambda)})} w(u_{\lambda(\mu)},u_{\kappa(\lambda)})\\
&+\alpha _1\alpha _2\alpha _3 \sum_{u_\kappa \in N(A)}\sum_{u_{\kappa(\lambda)} \in N(u_\kappa)} w(u_{\kappa(\lambda)},u_{\kappa})+\alpha _2\alpha _3\sum_{u_\kappa\in N(A)}w(u_\kappa,A)
\end{split}
\end{equation*}
As the weight is fixed, so $x_A^{(S')(4)}$ is mainly related to the number of neighbors, and the formula can be solved as $\alpha _2 \alpha _3(\alpha _1^3\mathcal{N}_A^{(S')(4)} +\alpha _1^2\mathcal{N}_A^{(S')(3)} + \alpha _1^1\mathcal{N}_A^{(S')(2)} + \alpha _1^0\mathcal{N}_A^{(S')(1)})$.
The representation in the network is approximately the sum of the number of neighbors of all nodes within four hops of node A.

Finally, we can get Eq.~\ref{eq:stage1}.

\noindent\textbf{Second stage.} Secondly, we study the right side of Eq.~\ref{eq:stage1}, which can be further derived as:
\begin{equation*}\small
\begin{split}
&|\beta''^T_1 \beta_3\alpha _2\alpha _3\sum_{i=1}^4\alpha _1^{i-1}[(\mathcal{N}_A^{(S')(i)}-\mathcal{N}_A^{(S)(i)})-(\mathcal{N}^{(S')(i)}_B-\mathcal{N}_B^{(S)(i)})]|\\
\le&\sum_{i=1}^4|\beta''^T_1 \beta_3\alpha _2\alpha _3\alpha _1^{i-1}[(\mathcal{N}_A^{(S')(i)}-\mathcal{N}_A^{(S)(i)})-(\mathcal{N}^{(S')(i)}_B-\mathcal{N}_B^{(S)(i)})]|
\end{split}
\end{equation*}
For ease of discussion, we denote the right part for $i=1$ to $4$ as $\Sigma_1$ to $\Sigma_4$, respectively. For the case $i=1$, it is not hard to see that only when the new seed, say $s$, appears as either $v_a$'s or $v_b$'s (but not both's) instant neighbor, $[(\mathcal{N}_A^{(S')(i)}-\mathcal{N}_A^{(S)(i)})-(\mathcal{N}^{(S')(i)}_B-\mathcal{N}_B^{(S)(i)})]$ is not zero. Notably, if $s$ is both $v_a$'s and $v_b$'s neighbor at the same time, the value within the square brackets will also be zero. Therefore,
\begin{equation*}\small
\begin{split}
E[\Sigma_1]=&(P[s\in N(v_a)^{(1)}, s\notin N(v_b)^{(1)}]+P[s\in N(v_b)^{(1)},s\notin N(v_a)^{(1)}])\\
&\beta''^T_5 \beta_3\alpha _2\alpha _3x_s\\
=&2P[s\in N(v_a)^{(1)},s\notin N(v_b)^{(1)}]Q(s)\\
\le& \frac{2|\overline{N(v)}|(n-|\overline{N(v)}|)}{n^2} \quad(\mbox{as } Q(s) \mbox{ has been normalized})\\
<& \frac{2|\overline{N(v)}|}{n^2}
\end{split}
\end{equation*}
where $N(v_a)^{(i)}$ denotes the set of nodes that are $i$-hop away from $v_a$, \eg $N(v_a)^{(1)}$ is in fact $N(v_a)$.

Then we carry on with $\Sigma_2$ when $i=2$. If $s$ is either $v_a$ or $v_b$'s 2-hop neighbor, but not both, $\Sigma_2$ appears to be not zero. Similar with $i=1$, the probability for this case is $\frac{2|\overline{N(v)}|^2(n-|\overline{N(v)}|^2)}{n^2}$. Besides, if $s$ appears to be the instant neighbor of $v_a$ or $v_b$, but not both, $\Sigma_2$ also appears positive. The probability for this case is the same with $i=1$, namely $\frac{2|\overline{N(v)}|(n-|\overline{N(v)}|)}{n^2}$. Then, the expectation of $\Sigma_2$ is as follows.
\begin{equation*}\small
\begin{split}
E[\Sigma_2]=&2P[s\in N(v_a)^{(2)},s\notin N(v_b)^{(2)}]Q(s)\\
+&2P[s\in N(v_a)^{(1)},s\notin N(v_b)^{(1)}]\sum_{v\in N(s)^{(1)}} Q(v)\\
=&\frac{2|\overline{N(v)}|^2(n-|\overline{N(v)}|^2)}{n^2}Q(s)+\frac{2|\overline{N(v)}|(n-|\overline{N(v)}|)}{n^2}\sum_{v\in N(s)^{(1)}} Q(v)\\
\le& \frac{2|\overline{N(v)}|^2(n-|\overline{N(v)}|^2)}{n^2}+\frac{2|\overline{N(v)}|^2(n-|\overline{N(v)}|)}{n^2}\\
<& \frac{4|\overline{N(v)}|^2}{n^2}.
\end{split}
\end{equation*}

Following the same way, we can derive $E[\Sigma_3]<\frac{6|\overline{N(v)}|^3}{n^2}$ and $E[\Sigma_4]<\frac{8|\overline{N(v)}|^4}{n^2}$.

In fact, it is obvious that $E[|(Q_A-Q_B)-(Q'_A-Q'_B)|]\le\sum_{i=1}^IE[\Sigma_i]$ and $\forall I<J, \sum_{i=1}^IE[\Sigma_i]<\sum_{i=1}^JE[\Sigma_i]$. That is, $\forall I<4$, $E[|(Q_A-Q_B)-(Q'_A-Q'_B)|]<\sum_{i=1}^4E[\Sigma_i]$.

Finally, we can prove that $\forall I<5$, $E[|(Q_A-Q_B)-(Q'_A-Q'_B)|]<\sum_{i=1}^4\frac{2i|\overline{N(v)}|^i}{n^2}$.
\vspace{0ex}

\eat{
\section*{References}
Please number citations consecutively within brackets \cite{b1}. The
sentence punctuation follows the bracket \cite{b2}. Refer simply to the reference
number, as in \cite{b3}---do not use ``Ref. \cite{b3}'' or ``reference \cite{b3}'' except at
the beginning of a sentence: ``Reference \cite{b3} was the first $\ldots$''

Number footnotes separately in superscripts. Place the actual footnote at
the bottom of the column in which it was cited. Do not put footnotes in the
abstract or reference list. Use letters for table footnotes.

Unless there are six authors or more give all authors' names; do not use
``et al.''. Papers that have not been published, even if they have been
submitted for publication, should be cited as ``unpublished'' \cite{b4}. Papers
that have been accepted for publication should be cited as ``in press'' \cite{b5}.
Capitalize only the first word in a paper title, except for proper nouns and
element symbols.

For papers published in translation journals, please give the English
citation first, followed by the original foreign-language citation \cite{b6}.

}
\balance

\end{document}